\documentclass[american,aps,prl,reprint]{revtex4-2}
\usepackage[T1]{fontenc}
\usepackage[latin9]{inputenc}
\usepackage{babel}
\usepackage{amsmath}
\usepackage{amsthm}
\usepackage{amssymb}
\usepackage{graphicx}
\usepackage[unicode=true,pdfusetitle,
 bookmarks=true,bookmarksnumbered=false,bookmarksopen=false,
 breaklinks=false,pdfborder={0 0 1},backref=false,colorlinks=false]
 {hyperref}
\hypersetup{
 pdfborderstyle=,colorlinks,allcolors=magenta}

\makeatletter
\usepackage{babel}
\usepackage{xcolor}
\usepackage{times}
\usepackage{txfonts}
\usepackage{braket}
\usepackage{colortbl}
\usepackage{bbm}
\theoremstyle{plain}

\newcommand{\id}{\mathbbm{1}}

\providecommand{\theoremname}{Theorem}

\makeatother

\begin{document}
\title{The power of noisy quantum states and the advantage of resource dilution}
\author{Marek Miller}
\author{Manfredi Scalici}
\author{Marco Fellous Asiani}
\author{Alexander Streltsov}
\email{a.streltsov@cent.uw.edu.pl}

\affiliation{Centre for Quantum Optical Technologies, Centre of New Technologies,
University of Warsaw, Banacha 2c, 02-097 Warsaw, Poland}
\begin{abstract}
Entanglement distillation allows to convert noisy quantum states into
singlets, which can in turn be used for various quantum technological
tasks, such as quantum teleportation and quantum key distribution.
Entanglement dilution is the inverse process: singlets are converted
into quantum states with less entanglement. While the usefulness of
distillation is apparent, practical applications of entanglement dilution
are less obvious. Here, we show that entanglement dilution can increase
the resilience of shared quantum states to local noise. The increased
resilience is observed even if diluting singlets into states with
arbitrarily little entanglement. We extend our analysis to other quantum
resource theories, such as quantum coherence, quantum thermodynamics,
and purity. For these resource theories, we demonstrate that diluting
pure quantum states into noisy ones can be advantageous for protecting
the system from noise. Our results demonstrate the usefulness of quantum
resource dilution, and provide a rare example for an advantage of
noisy quantum states over pure states in quantum information processing. 
\end{abstract}
\maketitle

As has been realized in the early days of quantum information theory,
two remote parties sharing a pair of entangled particles can perform
information processing tasks which are not possible in classical physics~\cite{HorodeckiRevModPhys.81.865}.
An important example of that is quantum key distribution~\cite{EkertPhysRevLett.67.661},
allowing the parties to establish a provably secure key. Typically,
these tasks employ singlets, highly entangled states of two quantum
bits. If the quantum states shared by the remote parties are noisy,
it is still possible to perform tasks based on singlets by applying
entanglement distillation~\cite{BennettPhysRevLett.76.722,BennettPhysRevA.53.2046}.
This procedure allows us to extract singlets from a large number of
copies of a noisy state, additionally making use of local operations
and classical communication (LOCC) between the remote parties. Quantum
states which can be converted into singlets in this way are called
\emph{distillable}. Since most quantum information processing tasks
are based on singlets, this makes all distillable states also useful
for these tasks. However, not all entangled states are distillable,
a phenomenon known as bound entanglement~\cite{HorodeckiPhysRevLett.80.5239}.

Conversely, it is possible to dilute singlets into quantum states
with less entanglement~\cite{BennettPhysRevA.53.2046}. For pure
entangled states, optimal distillation and dilution procedures are
known in the limit where a large number of copies of the state is
available~\cite{BennettPhysRevA.53.2046}. Two remote parties, Alice
and Bob, sharing a large number of copies of a pure entangled state
$\ket{\psi}^{AB}$ can distill them into singlets with the maximal
rate $S(\psi^{A})$, where $\psi^{A}=\mathrm{Tr}_{B}[\psi^{AB}]$
is the reduced state of Alice, $S(\rho)=-\mathrm{Tr}[\rho\log_{2}\rho]$
is the von Neumann entropy, and $\psi^{AB} = \ket{\psi}\!\bra{\psi}^{AB}$. The maximal rate for diluting singlets
into $\ket{\psi}^{AB}$ is given by $1/S(\psi^{A})$. For pure entangled
states the distillation and dilution procedures are reversible, which
means that in the asymptotic limit it is possible to distill $\ket{\psi}^{AB}$
into singlets and dilute them back into $\ket{\psi}^{AB}$ in a lossless
way~\cite{BennettPhysRevA.53.2046}.

\begin{figure*}
\includegraphics[width=0.6\paperwidth]{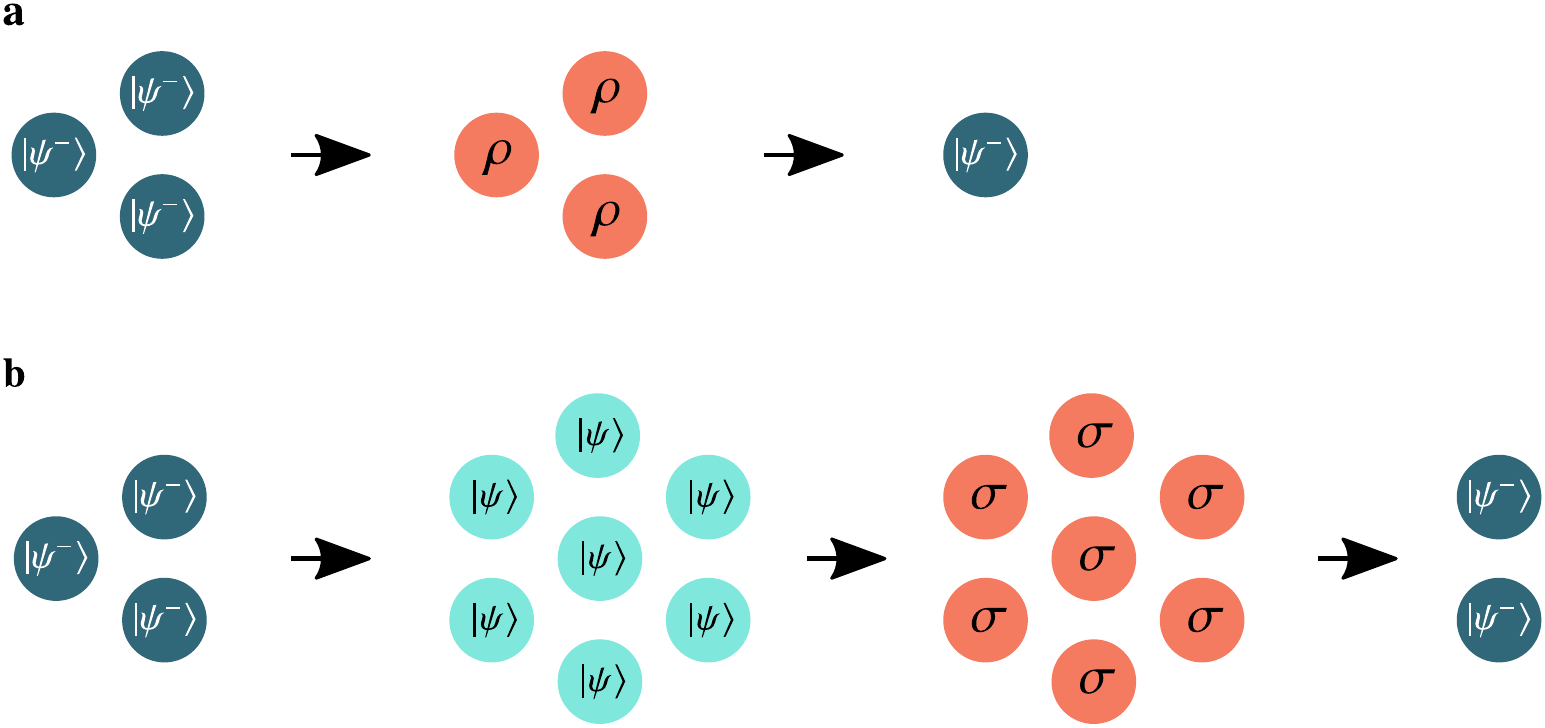}

\caption{Applying entanglement dilution to reduce the loss of entanglement
under local noise. Figure \textbf{a}) shows the setup without dilution:
a singlet $\ket{\psi^{-}}$ is subject to local noise on Bob's side,
resulting in the state $\rho=\openone\otimes\Lambda[\psi^{-}]$. In
our example, Alice and Bob can distill $\rho$ into singlets at rate
$1/3$. In figure \textbf{b}), Alice and Bob first dilute their singlets
into weakly entangled states $\ket{\psi}$. Each of these states undergoes
the same local noise as in figure \textbf{a}), resulting in the states
$\sigma=\openone\otimes\Lambda[\psi]$, which can then again be distilled
into singlets. The overall singlet rate is $2/3$, showing an improvement
over the setup in figure \textbf{a}).}
\label{fig:DilutionGeneral} 
\end{figure*}

While the dilution procedure is possible in principle, it is reasonable
to believe that in practice it is never advantageous to degrade singlets
into weakly entangled states. As we will see in this article, this
intuition is not correct: there exist quantum information processing
tasks where entanglement dilution is essential, even if the diluted
states contain arbitrarily little entanglement. Distillation and dilution
is not limited to entanglement, and has also been considered in general
quantum resource theories~\cite{ChitambarRevModPhys.91.025001}.
The basis of any quantum resource theory is the definition of free
states and free operations, corresponding to states and transformations
which can be created or performed at no cost within reasonable physical
constraints. Important examples are the resource theories of quantum
coherence~\cite{StreltsovRevModPhys.89.041003}, thermodynamics~\cite{Goold_2016},
and purity~\cite{HorodeckiPhysRevA.67.062104}. As we will see, resource
dilution provides an advantage in these quantum resource theories
as well.

\medskip
\textbf{\emph{Reducing entanglement loss under local noise. }}
Consider two remote parties, Alice and Bob, who share $n$ singlets $\ket{\psi^{-}}=(\ket{01}-\ket{10})/\sqrt{2}$.
We assume that Bob's quantum memory is not perfect, each qubit undergoing
local noise $\Lambda$. After the action of the noise, Alice and Bob
end up with $n$ copies of the noisy state $\rho=\openone\otimes\Lambda[\psi^{-}]$.
For large $n$, they can distill the states $\rho$ into $nE_{\mathrm{d}}(\rho)$
singlets, where $E_{\mathrm{d}}$ is the distillable entanglement~\cite{Plenioquant-ph/0504163,HorodeckiRevModPhys.81.865}. Since Alice and Bob started with $n$
singlets, $n[1-E_{\mathrm{d}}(\rho)]$ is the number of singlets lost
due to the imperfections of Bob's quantum memory.

As we will now show, Alice and Bob can reduce the loss of entanglement by diluting their singlets into states with less entanglement, see also Fig.~\ref{fig:DilutionGeneral}. By using LOCC,
Alice and Bob can dilute their $n$ singlets into $n/S(\psi^{A})$
copies of a weakly entangled state $\ket{\psi}$. We assume that this
dilution procedure can be achieved before the action of the noise.
Note that the number of diluted states $\ket{\psi}$ is larger than
the number of singlets $n$, and each of the additional qubits of
Bob is also subject to the same noise $\Lambda$, see Fig.~\ref{fig:DilutionGeneral}.
After the action of the noise, Alice and Bob end up sharing $n/S(\psi^{A})$
copies of the state $\sigma=\openone\otimes\Lambda[\psi]$, which
they can distill into singlets at rate $E_{\mathrm{d}}(\sigma)$.
Overall, in the limit of large $n$, Alice and Bob can obtain $nE_{\mathrm{d}}(\openone\otimes\Lambda[\psi])/S(\psi^{A})$
singlets using the dilution procedure.

\begin{figure}[b]
\includegraphics[width=1\columnwidth]{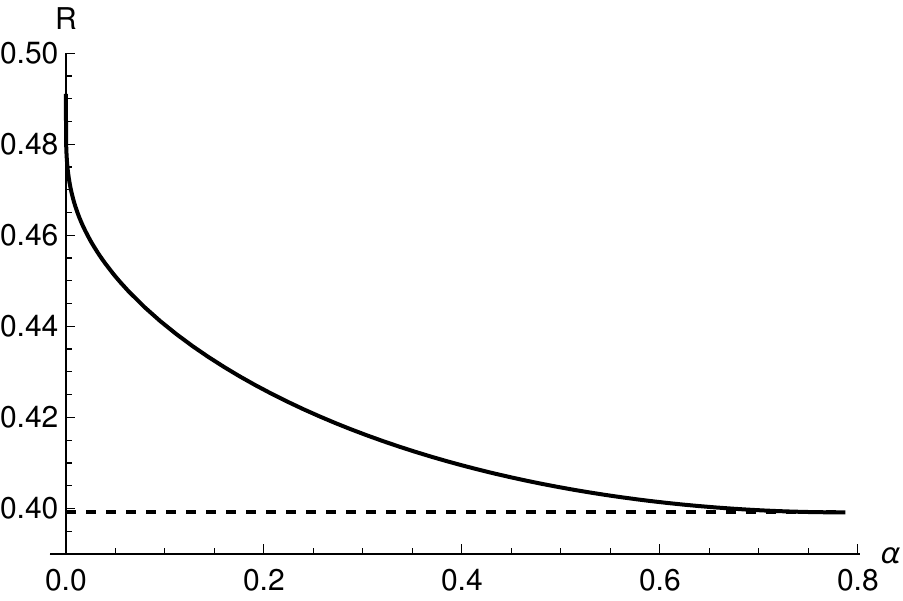}

\caption{Reducing entanglement loss under local phase damping by diluting into
pure states $\ket{\psi}=\cos\alpha\ket{00}+\sin\alpha\ket{11}$. Solid
curve shows $R=E_{\mathrm{d}}(\openone\otimes\Lambda[\psi])/S(\psi^{A})$
as a function of $\alpha$ for noise parameter $\lambda=1/2$. This
corresponds to the singlet rate achievable via dilution into $\ket{\psi}$.
Dashed line shows the corresponding singlet rate $E_{\mathrm{d}}(\openone\otimes\Lambda[\psi^{-}])$
if no dilution is applied. Maximal performance is achieved in the
limit $\alpha\rightarrow0$.}
\label{fig:Entanglement} 
\end{figure}

From the above discussion, it is clear that the dilution provides
an advantage whenever the inequality 
\begin{equation}
\frac{E_{\mathrm{d}}(\openone\otimes\Lambda[\psi])}{S(\psi^{A})}>E_{\mathrm{d}}(\openone\otimes\Lambda[\psi^{-}])\label{eq:DilutionPure}
\end{equation}
holds for some state $\ket{\psi}$. As we will now see, the dilution
procedure can indeed provide an advantage, even if the diluted states
$\ket{\psi}$ exhibit arbitrarily little entanglement. Suppose for example that Bob's qubits are subject to phase damping described by $\Lambda[\rho]=K_{0}\rho K_{0}^{\dagger}+K_{1}\rho K_{1}^{\dagger}$,
with Kraus operators 
\begin{equation}
K_{0}=\left(\begin{array}{cc}
1 & 0\\
0 & \sqrt{1-\lambda}
\end{array}\right),\,\,\,\,\,\,K_{1}=\left(\begin{array}{cc}
0 & 0\\
0 & \sqrt{\lambda}
\end{array}\right),\label{eq:PhaseDamping}
\end{equation}
and $0<\lambda<1$. We consider a situation when Alice and Bob dilute
their singlets into pure states $\ket{\psi}=\cos\alpha\ket{00}+\sin\alpha\ket{11}$.
In Fig.~\ref{fig:Entanglement}, we show both sides of the inequality~(\ref{eq:DilutionPure})
for $\lambda=1/2$ as a function of $\alpha$, see Supplemental Material for a more detailed analysis. We see that diluting the singlets provides
an advantage for all $\alpha$ in the range $0\leq\alpha < \pi/4$.
Moreover, the performance of dilution increases with decreasing $\alpha$,
reaching its maximal value for $\alpha\rightarrow0$. This behavior
is surprising, as for $\alpha=0$ the state is not entangled. 

\medskip
\textbf{\emph{Reducing the loss of coherence. }} We will now investigate
the usefulness of resource dilution for preserving quantum coherence.
Here, we assume that an initial collection of $n$ qubits in a maximally
coherent state $\ket{+}=(\ket{0}+\ket{1})\sqrt{2}$ undergoes a local
decoherence process $\Lambda$ (which we will specify later), leading
to $n$ copies of a final state $\rho=\Lambda[\ket{+}\!\bra{+}]$.
Similar to entanglement, we will now see that the loss of coherence
can be reduced by diluting the maximally coherent states into weakly
coherent ones.

In the following, we will focus on the resource theory of coherence
based on maximally incoherent operations~\cite{Abergquant-ph/0612146,StreltsovRevModPhys.89.041003}.
We take the free states to be diagonal in the reference basis $\{\ket{i}\}$,
and the free operations to be all operations which do not create coherence
in the reference basis. This is the largest set of operations which
is compatible with any reasonable resource theory of quantum coherence,
we refer to the Supplemental Material for more details.

After each of the qubits undergoes the decoherence process, in the
limit of large $n$ it is possible to distill the resulting states
$\rho$ into maximally coherent states at rate $C(\rho)=S(\Delta[\rho])-S(\rho)$~\cite{WinterPhysRevLett.116.120404}
with $\Delta[\rho]=\sum_{i}\braket{i|\rho|i}\ket{i}\!\bra{i}$. Similar
to entanglement dilution, it is possible to perform coherence dilution,
converting the maximally coherent states into weakly coherent states
$\mu$ at rate $1/C(\mu)$. Letting these states undergo the decoherence
process $\Lambda$, the overall rate of maximally coherent states
obtainable after the action of the noise is given by $C(\Lambda[\mu])/C(\mu)$.
As we will see in the following, dilution is useful for protecting
a system from a decoherence process. Moreover, we will see that for
some types of noise it is advantageous to dilute the states $\ket{+}$
into mixed states with little coherence.

\begin{figure}
\includegraphics[width=1\columnwidth]{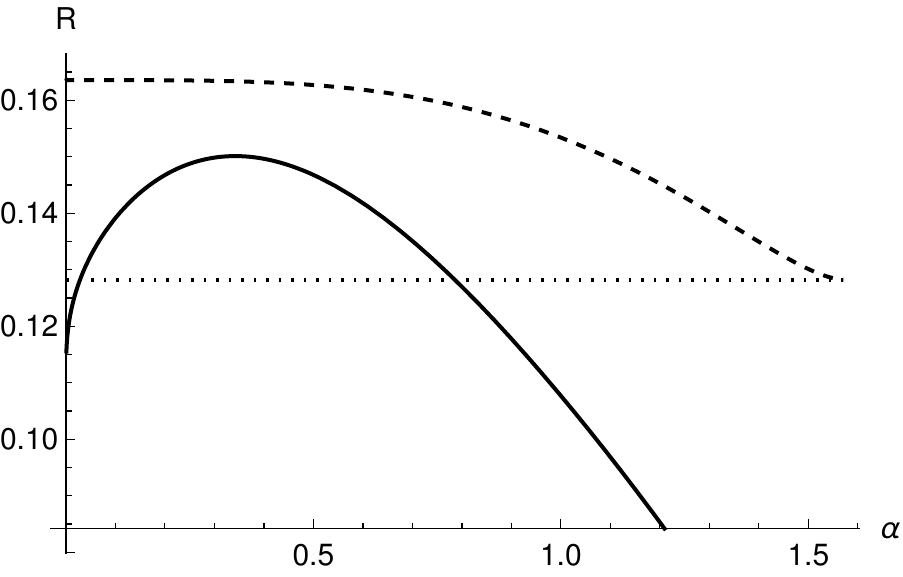}

\caption{Reducing the loss of coherence by dilution for single-qubit amplitude damping noise. We show the rate $R=C(\Lambda[\mu])/C(\mu)$
by dilution into pure states $\cos\alpha\ket{0}+\sin\alpha\ket{1}$
(solid curve) and mixed states $\sin^{2}\alpha\ket{+}\!\bra{+}+\cos^{2}\alpha\openone/2$
(dashed curve) for $\gamma=0.9$ as a function of $\alpha$. Dotted line
shows the coherence rate $R=C(\Lambda[\ket{+}\!\bra{+}])\approx0.13$
achievable without dilution.}
\label{fig:Coherence} 
\end{figure}

As an example demonstrating this effect, consider the single-qubit amplitude damping, which is represented by the Kraus operators
\begin{equation}
K_{0}=\left(\begin{array}{cc}
1 & 0\\
0 & \sqrt{1-\gamma}
\end{array}\right),\,\,\,\,\,\,K_{1}=\left(\begin{array}{cc}
0 & \sqrt{\gamma}\\
0 & 0
\end{array}\right).\label{AmpDamp}
\end{equation}
In Fig.~\ref{fig:Coherence}, we show the final rate of maximally
coherent states achievable without dilution, with dilution into pure
qubit states of the form $\ket{\psi}=\cos\alpha\ket{0}+\sin\alpha\ket{1}$,
and into mixed states of the form $\mu=\sin^{2}\alpha\ket{+}\!\bra{+}+\cos^{2}\alpha\openone/2$
for $\gamma=0.9$. As we see from Fig.~\ref{fig:Coherence}, by diluting
into pure qubit states it is possible to extract maximally coherent
states at an overall rate of 0.15, which is achieved for $\alpha\approx0.34$.
This is also the maximal possible rate achievable by dilution into
pure qubit states, see Supplemental Material for more details. In
contrast to this, diluting into the mixed state $\mu$ achieves maximal
performance in the limit $\alpha\rightarrow0$, leading to an overall
coherence rate $\log(19)/18\approx0.16$. Noting that the state $\mu$
is maximally mixed in this limit, this result is highly counterintuitive,
as it means that the best performance is obtained by creating a large
number of states which are almost maximally mixed.

\medskip
\textbf{\emph{Reducing the loss of energy and coherence in quantum
thermodynamics. }}As we will now see, the ideas presented above are
also applicable to the resource theory of quantum thermodynamics.
Here, we consider a quantum system $S$ with Hamiltonian $H_{S}$,
and the corresponding Gibbs state $\gamma^{S}=e^{-\beta H_{S}}/\mathrm{Tr}[e^{-\beta H_{S}}]$
at the inverse temperature $\beta=1/kT$. The Gibbs state is the free
state of the resource theory of quantum thermodynamics, and the free
transformations are known as thermal operations~\cite{Janzing2000}.
These are transformation of the system which can be implemented by
coupling the system to a thermal bath with Hamiltonian $H_{B}$ and
applying an energy preserving unitary: $\Lambda[\rho^{S}]=\mathrm{Tr}_{B}[U(\rho^{S}\otimes\gamma^{B})U^{\dagger}]$,
where $[U,H_{S}+H_{B}]=0$. Thermal operations preserve the Gibbs
state and do not increase the Helmholtz free energy of the system~\cite{Horodecki2013,BrandaoPhysRevLett.111.250404}.

If $n$ copies of a quantum state $\rho$ are available, then in the
limit $n\rightarrow\infty$ by using thermal operations it is possible
to convert $\rho$ into a state $\sigma$ which is diagonal in the
eigenbasis of $H_{S}$ at rate~\cite{BrandaoPhysRevLett.111.250404}
\begin{equation}
R(\rho\rightarrow\sigma)=\frac{S(\rho||\gamma)}{S(\sigma||\gamma)}\label{eq:GibbsPreservingRate}
\end{equation}
with the quantum relative entropy $S(\rho||\gamma)=\mathrm{Tr}[\rho\log_{2}\rho]-\mathrm{Tr}[\rho\log_{2}\gamma]$.
If $\sigma$ is not diagonal in the energy eigenbasis, the conversion
is possible at the same rate by using thermal operations together
with a sublinear number of qubits with coherence~\cite{BrandaoPhysRevLett.111.250404}.
It is possible to relax the set of free transformation to be Gibbs-preserving,
i.e., making the only requirement that they leave the Gibbs state
invariant~\cite{Faist_2015}. In this case, asymptotic transformations
are also characterized by Eq.~(\ref{eq:GibbsPreservingRate}), as
follows from results in~\cite{WangPhysRevResearch.1.033170}. In
contrast to thermal operations, no additional coherence is required
in this setup.

In the following, we will focus on qubit systems with Hamiltonian
$H=E_{0}\ket{E_{0}}\!\bra{E_{0}}+E_{1}\ket{E_{1}}\!\bra{E_{1}}$ at
temperature $T$, where $\ket{E_{i}}$ are the eigenstates of the
Hamiltonian with eigenvalues $E_{i}$, and $E_{0}$ is the ground
state energy. Consider now $n$ qubits, initialized in the excited
state $\ket{E_{1}}$. We assume that each of the qubits is subject
to a \emph{thermal noise} $\Lambda$, i.e., noise which does not create
coherence in the eigenbasis of $H_{S}$ and does not increase the
Helmholtz free energy of the system. As we will now see, diluting
the excited state $\ket{E_{1}}$ into noisy states can provide an
advantage, making the overall $n$ qubit system more robust against
the action of thermal noise.

In analogy to our previous results for entanglement and coherence,
by diluting $n$ copies of the excited states into $\mu$ before the
action of the noise it is possible to obtain excited states at the
overall rate $S(\Lambda[\mu]||\gamma)/S(\mu||\gamma)$. Dilution of
the excited state into $\mu$ provides an advantage whenever 
\begin{equation}
\frac{S(\Lambda[\mu]||\gamma)}{S(\mu||\gamma)}>\frac{S(\Lambda[\ket{E_{1}}\!\bra{E_{1}}]||\gamma)}{S(\ket{E_{1}}\!\bra{E_{1}}||\gamma)}.
\end{equation}
As an example, consider now noise of the form 
\begin{equation}
\Lambda[\rho]=p\gamma+(1-p)\Delta[\rho],\label{eq:ThermalNoise}
\end{equation}
with $\Delta[\rho]=\sum_{i}\braket{E_{i}|\rho|E_{i}}\ket{E_{i}}\!\bra{E_{i}}$
and $0\leq p\leq1$. It is clear that the noise $\Lambda$ is thermal,
as it destroys all coherence eventually available in the state $\rho$
and does not increase the Helmholtz free energy of the system. For
this type of noise, optimal performance can be achieved by diluting
the excited state into a diagonal state: 
\begin{equation}
\mu=(1-q)\ket{E_{0}}\!\bra{E_{0}}+q\ket{E_{1}}\!\bra{E_{1}},\label{eq:ThermalOptimal}
\end{equation}
we refer to the Supplemental Material for more details.

So far we have focused on transformations between states which are
diagonal in the energy eigenbasis. We will now go one step further,
assuming that the initial state has coherence, i.e., $\ket{\psi}=\cos\alpha\ket{E_{0}}+\sin\alpha\ket{E_{1}}$.
Our goal now is to protect this state from decoherence in the eigenbasis
of the Hamiltonian, i.e., from thermal noise with Kraus operators
as given in Eq.~(\ref{eq:PhaseDamping}). Protecting $\ket{\psi}$
from decoherence can be achieved by transforming the state into some
diagonal state $\mu$ of the form~(\ref{eq:ThermalOptimal}). Note
that any diagonal state is invariant under decoherence. For large
number of copies of the initial state, the conversion into $\mu$
is possible via thermal operations at rate $S(\psi||\gamma)/S(\mu||\gamma)$~\cite{BrandaoPhysRevLett.111.250404},
where we assume that $\mu\neq\gamma$. Moreover, it is possible to
convert $\mu$ back into $\ket{\psi}$ via Gibbs-preserving operations
at rate $S(\mu||\gamma)/S(\psi||\gamma)$~\cite{WangPhysRevResearch.1.033170}.
For transformations via thermal operations, additionally a sublinear
number of qubits with coherence is required~\cite{BrandaoPhysRevLett.111.250404}.
The overall procedure is reversible in the asymptotic limit. Thus,
it is possible to completely protect a system from decoherence in
the eigenbasis of the Hamiltonian using Gibbs-preserving operations.
The same is true for thermal operations, if we have access to a sublinear
number of qubits with coherence.

\label{purityPreserve} Another resource theory which is closely related
to quantum thermodynamics is the resource theory of purity~\cite{HorodeckiPhysRevA.67.062104,GOUR20151}.
Here, the free state is the maximally mixed state $\openone/d$ and
the free operations are unital, i.e., $\Lambda[\openone/d]=\openone/d$.
Asymptotic transformation rates in this theory are given by Eq.~(\ref{eq:GibbsPreservingRate})
if we set $\gamma=\openone/d$~\cite{HorodeckiPhysRevA.67.062104}.
By the same arguments as above, we see that the rate for preserving
a pure state from unital noise $\Lambda$ by diluting it into a noisy
state $\mu$ is given by $S\left(\Lambda[\mu]||\openone/d\right)/S\left(\mu||\openone/d\right)$.

We will now focus on single-qubit settings. As an example, consider single-qubit
depolarizing noise $\Lambda[\rho]=p\openone/2+(1-p)\rho$. For protecting the system from this noise by diluting it into a state
$\mu$, it is enough to consider diagonal states of the form $\mu=(1-q)\ket{0}\!\bra{0}+q\ket{1}\!\bra{1}$. As we prove in the Supplemental Material, for any $0<p<1$ the rate is maximal in the limit $q\rightarrow1/2$.
Note that the state $\mu$ is maximally mixed in this limit, thus
the performance increases with increased level of dilution. As we discuss in the Supplemental Material, similar statement can be made for general depolarizing noise of a $d$-dimensional systems.

Our discussion on the resource theory of purity has so far focused
on single-qubit systems. Instead of diluting $n$ pure qubit
states into noisy states of a single qubit, one could transform them
into some correlated $k$-qubit states $\mu_{k}$. Can correlations in the diluted state $\mu_{k}$ increase the performance
of the procedure? As we discuss in the Supplemental Material, correlations
are not useful when the state $\mu_{k}$ is pure. This result shows
that diluting into noisy states provides an advantage even if correlations
are taken into account.

\medskip
\textbf{\emph{General quantum resource theories and strategies beyond
dilution.}} In a general quantum resource theory, asymptotic conversion
between states is determined by conversion rates $R(\rho\rightarrow\sigma)$,
we refer to the Supplemental Material for a formal definition. In
general, our goal is to preserve a noisy quantum system in the pure
state $\ket{\psi}$, where the noise is described by $\Lambda$, and
we assume that $\Lambda$ does not generate the resource under consideration.
If $n$ copies of the state $\ket{\psi}$ are subject to noise, then
in the limit $n\rightarrow\infty$ the resulting states $\Lambda[\psi]$
can be converted back into the original state $\ket{\psi}$ at rate
$R(\Lambda[\psi]\rightarrow\psi)$.

In analogy to our discussion for entanglement, coherence, and thermodynamics,
we assume that it is possible to dilute $n$ copies of the initial
state $\ket{\psi}$ into another state $\mu$ before the action of
the noise. For large $n$, this is possible at the rate $R(\psi\rightarrow\mu)$.
The resulting state $\mu$ is then subject to noise $\Lambda$. Since
our goal is to preserve the system in the initial state $\ket{\psi}$,
the state $\Lambda[\mu]$ is converted back into $\ket{\psi}$. The
overall rate of the conversion procedure (i.e. the number of final
states $\ket{\psi}$ per copy of the initial state) is given by $R(\psi\rightarrow\mu)\times R(\Lambda[\mu]\rightarrow\psi)$.
Diluting into the state $\mu$ provides an advantage whenever 
\begin{equation}
R(\psi\rightarrow\mu)\times R(\Lambda[\mu]\rightarrow\psi)>R(\Lambda[\psi]\rightarrow\psi).
\end{equation}
If the resource theory is reversible, then the overall
rate $R(\psi\rightarrow\mu)\times R(\Lambda[\mu]\rightarrow\psi)$
is the same for all resource states $\ket{\psi}$, we refer to the
Supplemental Material for more details.

For the resource theories of entanglement, coherence, quantum thermodynamics,
and purity we have seen that resource dilution provides an advantage
for protecting a quantum system from noise. However, dilution is not
the most general strategy. When it comes to preserving entanglement against local decoherence on Bob's side, Bob might as well apply a local quantum error correction scheme. In the Supplemental Material we report a detailed numerical comparison between the dilution procedure and quantum error correction for single-qubit Pauli noise. The numerical results suggest that at least in the setting considered here quantum error correction outperforms entanglement dilution, when it comes to preserving a large number of singlets. As we also demonstrate in the Supplemental Material, the dilution strategy is useful also for noise of larger dimension. It remains an open question how dilution compares to quantum error correction in this general case.

\medskip
\textbf{\emph{Conclusions.}} We have seen that diluting quantum resources
has practical applications for protecting a quantum system from noise.
This applies for entanglement, where diluting singlets into weakly
entangled states can make the system more resilient against the action
of local noise. Similar effects were found in the resource theories
of coherence, quantum thermodynamics, and purity. Even more, for these
resource theories we found that for some types of noise it is advantageous
to dilute the system into a noisy state. This result is highly counterintuitive,
as it demonstrates that noisy states can outperform pure states in
quantum information processing tasks.

Several intriguing questions are left open in this work. One such
open question concerns the role of noisy states in entanglement dilution.
While we have seen that diluting a singlet into a weakly entangled
pure state is advantageous for some types of local noise, it remains
unclear if dilution into a noisy entangled state can perform even
better. Moreover, our results suggest that resource dilution might
be useful for most quantum resource theories, and for most types of
resource non-generating noise. To prove these statements is left open
for future research.

We thank Ray Ganardi, Tulja Varun Kondra, Ludovico Lami, Carlo Marconi, Martin Plenio, Bartosz Regula, Anna Sanpera, and Andreas Winter for discussion. This work was supported by the ``Quantum Optical Technologies'' project, carried out within the International Research Agendas programme of the Foundation for Polish Science co-financed by the European Union under the European Regional Development Fund, the ``Quantum Coherence and Entanglement for Quantum Technology'' project, carried out within the First Team programme of the Foundation for Polish Science co-financed by the European Union under the European Regional Development Fund, and the National Science Centre, Poland, within the QuantERA II Programme (No 2021/03/Y/ST2/00178, acronym ExTRaQT) that has received funding from the European Union's Horizon 2020 research and innovation programme under Grant Agreement No 101017733.

\bibliography{literature}

\begin{thebibliography}{33}%
\makeatletter
\providecommand \@ifxundefined [1]{%
 \@ifx{#1\undefined}
}%
\providecommand \@ifnum [1]{%
 \ifnum #1\expandafter \@firstoftwo
 \else \expandafter \@secondoftwo
 \fi
}%
\providecommand \@ifx [1]{%
 \ifx #1\expandafter \@firstoftwo
 \else \expandafter \@secondoftwo
 \fi
}%
\providecommand \natexlab [1]{#1}%
\providecommand \enquote  [1]{``#1''}%
\providecommand \bibnamefont  [1]{#1}%
\providecommand \bibfnamefont [1]{#1}%
\providecommand \citenamefont [1]{#1}%
\providecommand \href@noop [0]{\@secondoftwo}%
\providecommand \href [0]{\begingroup \@sanitize@url \@href}%
\providecommand \@href[1]{\@@startlink{#1}\@@href}%
\providecommand \@@href[1]{\endgroup#1\@@endlink}%
\providecommand \@sanitize@url [0]{\catcode `\\12\catcode `\$12\catcode
  `\&12\catcode `\#12\catcode `\^12\catcode `\_12\catcode `\%12\relax}%
\providecommand \@@startlink[1]{}%
\providecommand \@@endlink[0]{}%
\providecommand \url  [0]{\begingroup\@sanitize@url \@url }%
\providecommand \@url [1]{\endgroup\@href {#1}{\urlprefix }}%
\providecommand \urlprefix  [0]{URL }%
\providecommand \Eprint [0]{\href }%
\providecommand \doibase [0]{https://doi.org/}%
\providecommand \selectlanguage [0]{\@gobble}%
\providecommand \bibinfo  [0]{\@secondoftwo}%
\providecommand \bibfield  [0]{\@secondoftwo}%
\providecommand \translation [1]{[#1]}%
\providecommand \BibitemOpen [0]{}%
\providecommand \bibitemStop [0]{}%
\providecommand \bibitemNoStop [0]{.\EOS\space}%
\providecommand \EOS [0]{\spacefactor3000\relax}%
\providecommand \BibitemShut  [1]{\csname bibitem#1\endcsname}%
\let\auto@bib@innerbib\@empty
\bibitem [{\citenamefont {Horodecki}\ \emph {et~al.}(2009)\citenamefont
  {Horodecki}, \citenamefont {Horodecki}, \citenamefont {Horodecki},\ and\
  \citenamefont {Horodecki}}]{HorodeckiRevModPhys.81.865}%
  \BibitemOpen
  \bibfield  {author} {\bibinfo {author} {\bibfnamefont {R.}~\bibnamefont
  {Horodecki}}, \bibinfo {author} {\bibfnamefont {P.}~\bibnamefont
  {Horodecki}}, \bibinfo {author} {\bibfnamefont {M.}~\bibnamefont
  {Horodecki}},\ and\ \bibinfo {author} {\bibfnamefont {K.}~\bibnamefont
  {Horodecki}},\ }\bibfield  {title} {\bibinfo {title} {Quantum entanglement},\
  }\href {https://doi.org/10.1103/RevModPhys.81.865} {\bibfield  {journal}
  {\bibinfo  {journal} {Rev. Mod. Phys.}\ }\textbf {\bibinfo {volume} {81}},\
  \bibinfo {pages} {865} (\bibinfo {year} {2009})}\BibitemShut {NoStop}%
\bibitem [{\citenamefont {Ekert}(1991)}]{EkertPhysRevLett.67.661}%
  \BibitemOpen
  \bibfield  {author} {\bibinfo {author} {\bibfnamefont {A.~K.}\ \bibnamefont
  {Ekert}},\ }\bibfield  {title} {\bibinfo {title} {{Quantum cryptography based
  on Bell's theorem}},\ }\href {https://doi.org/10.1103/PhysRevLett.67.661}
  {\bibfield  {journal} {\bibinfo  {journal} {Phys. Rev. Lett.}\ }\textbf
  {\bibinfo {volume} {67}},\ \bibinfo {pages} {661} (\bibinfo {year}
  {1991})}\BibitemShut {NoStop}%
\bibitem [{\citenamefont {Bennett}\ \emph
  {et~al.}(1996{\natexlab{a}})\citenamefont {Bennett}, \citenamefont
  {Brassard}, \citenamefont {Popescu}, \citenamefont {Schumacher},
  \citenamefont {Smolin},\ and\ \citenamefont
  {Wootters}}]{BennettPhysRevLett.76.722}%
  \BibitemOpen
  \bibfield  {author} {\bibinfo {author} {\bibfnamefont {C.~H.}\ \bibnamefont
  {Bennett}}, \bibinfo {author} {\bibfnamefont {G.}~\bibnamefont {Brassard}},
  \bibinfo {author} {\bibfnamefont {S.}~\bibnamefont {Popescu}}, \bibinfo
  {author} {\bibfnamefont {B.}~\bibnamefont {Schumacher}}, \bibinfo {author}
  {\bibfnamefont {J.~A.}\ \bibnamefont {Smolin}},\ and\ \bibinfo {author}
  {\bibfnamefont {W.~K.}\ \bibnamefont {Wootters}},\ }\bibfield  {title}
  {\bibinfo {title} {{Purification of Noisy Entanglement and Faithful
  Teleportation via Noisy Channels}},\ }\href
  {https://doi.org/10.1103/PhysRevLett.76.722} {\bibfield  {journal} {\bibinfo
  {journal} {Phys. Rev. Lett.}\ }\textbf {\bibinfo {volume} {76}},\ \bibinfo
  {pages} {722} (\bibinfo {year} {1996}{\natexlab{a}})}\BibitemShut {NoStop}%
\bibitem [{\citenamefont {Bennett}\ \emph
  {et~al.}(1996{\natexlab{b}})\citenamefont {Bennett}, \citenamefont
  {Bernstein}, \citenamefont {Popescu},\ and\ \citenamefont
  {Schumacher}}]{BennettPhysRevA.53.2046}%
  \BibitemOpen
  \bibfield  {author} {\bibinfo {author} {\bibfnamefont {C.~H.}\ \bibnamefont
  {Bennett}}, \bibinfo {author} {\bibfnamefont {H.~J.}\ \bibnamefont
  {Bernstein}}, \bibinfo {author} {\bibfnamefont {S.}~\bibnamefont {Popescu}},\
  and\ \bibinfo {author} {\bibfnamefont {B.}~\bibnamefont {Schumacher}},\
  }\bibfield  {title} {\bibinfo {title} {Concentrating partial entanglement by
  local operations},\ }\href {https://doi.org/10.1103/PhysRevA.53.2046}
  {\bibfield  {journal} {\bibinfo  {journal} {Phys. Rev. A}\ }\textbf {\bibinfo
  {volume} {53}},\ \bibinfo {pages} {2046} (\bibinfo {year}
  {1996}{\natexlab{b}})}\BibitemShut {NoStop}%
\bibitem [{\citenamefont {Horodecki}\ \emph {et~al.}(1998)\citenamefont
  {Horodecki}, \citenamefont {Horodecki},\ and\ \citenamefont
  {Horodecki}}]{HorodeckiPhysRevLett.80.5239}%
  \BibitemOpen
  \bibfield  {author} {\bibinfo {author} {\bibfnamefont {M.}~\bibnamefont
  {Horodecki}}, \bibinfo {author} {\bibfnamefont {P.}~\bibnamefont
  {Horodecki}},\ and\ \bibinfo {author} {\bibfnamefont {R.}~\bibnamefont
  {Horodecki}},\ }\bibfield  {title} {\bibinfo {title} {{Mixed-State
  Entanglement and Distillation: Is there a ``Bound'' Entanglement in
  Nature?}},\ }\href {https://doi.org/10.1103/PhysRevLett.80.5239} {\bibfield
  {journal} {\bibinfo  {journal} {Phys. Rev. Lett.}\ }\textbf {\bibinfo
  {volume} {80}},\ \bibinfo {pages} {5239} (\bibinfo {year}
  {1998})}\BibitemShut {NoStop}%
\bibitem [{\citenamefont {Chitambar}\ and\ \citenamefont
  {Gour}(2019)}]{ChitambarRevModPhys.91.025001}%
  \BibitemOpen
  \bibfield  {author} {\bibinfo {author} {\bibfnamefont {E.}~\bibnamefont
  {Chitambar}}\ and\ \bibinfo {author} {\bibfnamefont {G.}~\bibnamefont
  {Gour}},\ }\bibfield  {title} {\bibinfo {title} {Quantum resource theories},\
  }\href {https://doi.org/10.1103/RevModPhys.91.025001} {\bibfield  {journal}
  {\bibinfo  {journal} {Rev. Mod. Phys.}\ }\textbf {\bibinfo {volume} {91}},\
  \bibinfo {pages} {025001} (\bibinfo {year} {2019})}\BibitemShut {NoStop}%
\bibitem [{\citenamefont {Streltsov}\ \emph {et~al.}(2017)\citenamefont
  {Streltsov}, \citenamefont {Adesso},\ and\ \citenamefont
  {Plenio}}]{StreltsovRevModPhys.89.041003}%
  \BibitemOpen
  \bibfield  {author} {\bibinfo {author} {\bibfnamefont {A.}~\bibnamefont
  {Streltsov}}, \bibinfo {author} {\bibfnamefont {G.}~\bibnamefont {Adesso}},\
  and\ \bibinfo {author} {\bibfnamefont {M.~B.}\ \bibnamefont {Plenio}},\
  }\bibfield  {title} {\bibinfo {title} {Colloquium: Quantum coherence as a
  resource},\ }\href {https://doi.org/10.1103/RevModPhys.89.041003} {\bibfield
  {journal} {\bibinfo  {journal} {Rev. Mod. Phys.}\ }\textbf {\bibinfo {volume}
  {89}},\ \bibinfo {pages} {041003} (\bibinfo {year} {2017})}\BibitemShut
  {NoStop}%
\bibitem [{\citenamefont {Goold}\ \emph {et~al.}(2016)\citenamefont {Goold},
  \citenamefont {Huber}, \citenamefont {Riera}, \citenamefont {del Rio},\ and\
  \citenamefont {Skrzypczyk}}]{Goold_2016}%
  \BibitemOpen
  \bibfield  {author} {\bibinfo {author} {\bibfnamefont {J.}~\bibnamefont
  {Goold}}, \bibinfo {author} {\bibfnamefont {M.}~\bibnamefont {Huber}},
  \bibinfo {author} {\bibfnamefont {A.}~\bibnamefont {Riera}}, \bibinfo
  {author} {\bibfnamefont {L.}~\bibnamefont {del Rio}},\ and\ \bibinfo {author}
  {\bibfnamefont {P.}~\bibnamefont {Skrzypczyk}},\ }\bibfield  {title}
  {\bibinfo {title} {The role of quantum information in thermodynamics - a
  topical review},\ }\href {https://doi.org/10.1088/1751-8113/49/14/143001}
  {\bibfield  {journal} {\bibinfo  {journal} {J. Phys. A}\ }\textbf {\bibinfo
  {volume} {49}},\ \bibinfo {pages} {143001} (\bibinfo {year}
  {2016})}\BibitemShut {NoStop}%
\bibitem [{\citenamefont {Horodecki}\ \emph {et~al.}(2003)\citenamefont
  {Horodecki}, \citenamefont {Horodecki},\ and\ \citenamefont
  {Oppenheim}}]{HorodeckiPhysRevA.67.062104}%
  \BibitemOpen
  \bibfield  {author} {\bibinfo {author} {\bibfnamefont {M.}~\bibnamefont
  {Horodecki}}, \bibinfo {author} {\bibfnamefont {P.}~\bibnamefont
  {Horodecki}},\ and\ \bibinfo {author} {\bibfnamefont {J.}~\bibnamefont
  {Oppenheim}},\ }\bibfield  {title} {\bibinfo {title} {Reversible
  transformations from pure to mixed states and the unique measure of
  information},\ }\href {https://doi.org/10.1103/PhysRevA.67.062104} {\bibfield
   {journal} {\bibinfo  {journal} {Phys. Rev. A}\ }\textbf {\bibinfo {volume}
  {67}},\ \bibinfo {pages} {062104} (\bibinfo {year} {2003})}\BibitemShut
  {NoStop}%
\bibitem [{\citenamefont {Plenio}\ and\ \citenamefont
  {Virmani}(2007)}]{Plenioquant-ph/0504163}%
  \BibitemOpen
  \bibfield  {author} {\bibinfo {author} {\bibfnamefont {M.~B.}\ \bibnamefont
  {Plenio}}\ and\ \bibinfo {author} {\bibfnamefont {S.}~\bibnamefont
  {Virmani}},\ }\bibfield  {title} {\bibinfo {title} {An introduction to
  entanglement measures},\ }\href@noop {} {\bibfield  {journal} {\bibinfo
  {journal} {Quant. Inf. Comput.}\ }\textbf {\bibinfo {volume} {7}},\ \bibinfo
  {pages} {1} (\bibinfo {year} {2007})},\ \Eprint
  {https://arxiv.org/abs/arXiv:quant-ph/0504163} {arXiv:quant-ph/0504163}
  \BibitemShut {NoStop}%
\bibitem [{\citenamefont {Aberg}(2006)}]{Abergquant-ph/0612146}%
  \BibitemOpen
  \bibfield  {author} {\bibinfo {author} {\bibfnamefont {J.}~\bibnamefont
  {Aberg}},\ }\bibfield  {title} {\bibinfo {title} {{Quantifying
  Superposition}},\ }\href {https://arxiv.org/abs/quant-ph/0612146} {\bibfield
  {journal} {\bibinfo  {journal} {arXiv:quant-ph/0612146}\ } (\bibinfo {year}
  {2006})}\BibitemShut {NoStop}%
\bibitem [{\citenamefont {Winter}\ and\ \citenamefont
  {Yang}(2016)}]{WinterPhysRevLett.116.120404}%
  \BibitemOpen
  \bibfield  {author} {\bibinfo {author} {\bibfnamefont {A.}~\bibnamefont
  {Winter}}\ and\ \bibinfo {author} {\bibfnamefont {D.}~\bibnamefont {Yang}},\
  }\bibfield  {title} {\bibinfo {title} {{Operational Resource Theory of
  Coherence}},\ }\href {https://doi.org/10.1103/PhysRevLett.116.120404}
  {\bibfield  {journal} {\bibinfo  {journal} {Phys. Rev. Lett.}\ }\textbf
  {\bibinfo {volume} {116}},\ \bibinfo {pages} {120404} (\bibinfo {year}
  {2016})}\BibitemShut {NoStop}%
\bibitem [{\citenamefont {Janzing}\ \emph {et~al.}(2000)\citenamefont
  {Janzing}, \citenamefont {Wocjan}, \citenamefont {Zeier}, \citenamefont
  {Geiss},\ and\ \citenamefont {Beth}}]{Janzing2000}%
  \BibitemOpen
  \bibfield  {author} {\bibinfo {author} {\bibfnamefont {D.}~\bibnamefont
  {Janzing}}, \bibinfo {author} {\bibfnamefont {P.}~\bibnamefont {Wocjan}},
  \bibinfo {author} {\bibfnamefont {R.}~\bibnamefont {Zeier}}, \bibinfo
  {author} {\bibfnamefont {R.}~\bibnamefont {Geiss}},\ and\ \bibinfo {author}
  {\bibfnamefont {T.}~\bibnamefont {Beth}},\ }\bibfield  {title} {\bibinfo
  {title} {{Thermodynamic Cost of Reliability and Low Temperatures: Tightening
  Landauer's Principle and the Second Law}},\ }\href
  {https://doi.org/10.1023/A:1026422630734} {\bibfield  {journal} {\bibinfo
  {journal} {International Journal of Theoretical Physics}\ }\textbf {\bibinfo
  {volume} {39}},\ \bibinfo {pages} {2717} (\bibinfo {year}
  {2000})}\BibitemShut {NoStop}%
\bibitem [{\citenamefont {Horodecki}\ and\ \citenamefont
  {Oppenheim}(2013{\natexlab{a}})}]{Horodecki2013}%
  \BibitemOpen
  \bibfield  {author} {\bibinfo {author} {\bibfnamefont {M.}~\bibnamefont
  {Horodecki}}\ and\ \bibinfo {author} {\bibfnamefont {J.}~\bibnamefont
  {Oppenheim}},\ }\bibfield  {title} {\bibinfo {title} {Fundamental limitations
  for quantum and nanoscale thermodynamics},\ }\href
  {https://doi.org/10.1038/ncomms3059} {\bibfield  {journal} {\bibinfo
  {journal} {Nature Communications}\ }\textbf {\bibinfo {volume} {4}},\
  \bibinfo {pages} {2059} (\bibinfo {year} {2013}{\natexlab{a}})}\BibitemShut
  {NoStop}%
\bibitem [{\citenamefont {Brand\~ao}\ \emph {et~al.}(2013)\citenamefont
  {Brand\~ao}, \citenamefont {Horodecki}, \citenamefont {Oppenheim},
  \citenamefont {Renes},\ and\ \citenamefont
  {Spekkens}}]{BrandaoPhysRevLett.111.250404}%
  \BibitemOpen
  \bibfield  {author} {\bibinfo {author} {\bibfnamefont {F.~G. S.~L.}\
  \bibnamefont {Brand\~ao}}, \bibinfo {author} {\bibfnamefont {M.}~\bibnamefont
  {Horodecki}}, \bibinfo {author} {\bibfnamefont {J.}~\bibnamefont
  {Oppenheim}}, \bibinfo {author} {\bibfnamefont {J.~M.}\ \bibnamefont
  {Renes}},\ and\ \bibinfo {author} {\bibfnamefont {R.~W.}\ \bibnamefont
  {Spekkens}},\ }\bibfield  {title} {\bibinfo {title} {{Resource Theory of
  Quantum States Out of Thermal Equilibrium}},\ }\href
  {https://doi.org/10.1103/PhysRevLett.111.250404} {\bibfield  {journal}
  {\bibinfo  {journal} {Phys. Rev. Lett.}\ }\textbf {\bibinfo {volume} {111}},\
  \bibinfo {pages} {250404} (\bibinfo {year} {2013})}\BibitemShut {NoStop}%
\bibitem [{\citenamefont {Faist}\ \emph {et~al.}(2015)\citenamefont {Faist},
  \citenamefont {Oppenheim},\ and\ \citenamefont {Renner}}]{Faist_2015}%
  \BibitemOpen
  \bibfield  {author} {\bibinfo {author} {\bibfnamefont {P.}~\bibnamefont
  {Faist}}, \bibinfo {author} {\bibfnamefont {J.}~\bibnamefont {Oppenheim}},\
  and\ \bibinfo {author} {\bibfnamefont {R.}~\bibnamefont {Renner}},\
  }\bibfield  {title} {\bibinfo {title} {Gibbs-preserving maps outperform
  thermal operations in the quantum regime},\ }\href
  {https://doi.org/10.1088/1367-2630/17/4/043003} {\bibfield  {journal}
  {\bibinfo  {journal} {New J. Phys.}\ }\textbf {\bibinfo {volume} {17}},\
  \bibinfo {pages} {043003} (\bibinfo {year} {2015})}\BibitemShut {NoStop}%
\bibitem [{\citenamefont {Wang}\ and\ \citenamefont
  {Wilde}(2019)}]{WangPhysRevResearch.1.033170}%
  \BibitemOpen
  \bibfield  {author} {\bibinfo {author} {\bibfnamefont {X.}~\bibnamefont
  {Wang}}\ and\ \bibinfo {author} {\bibfnamefont {M.~M.}\ \bibnamefont
  {Wilde}},\ }\bibfield  {title} {\bibinfo {title} {Resource theory of
  asymmetric distinguishability},\ }\href
  {https://doi.org/10.1103/PhysRevResearch.1.033170} {\bibfield  {journal}
  {\bibinfo  {journal} {Phys. Rev. Research}\ }\textbf {\bibinfo {volume}
  {1}},\ \bibinfo {pages} {033170} (\bibinfo {year} {2019})}\BibitemShut
  {NoStop}%
\bibitem [{\citenamefont {Gour}\ \emph {et~al.}(2015)\citenamefont {Gour},
  \citenamefont {M\"uller}, \citenamefont {Narasimhachar}, \citenamefont
  {Spekkens},\ and\ \citenamefont {{Yunger Halpern}}}]{GOUR20151}%
  \BibitemOpen
  \bibfield  {author} {\bibinfo {author} {\bibfnamefont {G.}~\bibnamefont
  {Gour}}, \bibinfo {author} {\bibfnamefont {M.~P.}\ \bibnamefont {M\"uller}},
  \bibinfo {author} {\bibfnamefont {V.}~\bibnamefont {Narasimhachar}}, \bibinfo
  {author} {\bibfnamefont {R.~W.}\ \bibnamefont {Spekkens}},\ and\ \bibinfo
  {author} {\bibfnamefont {N.}~\bibnamefont {{Yunger Halpern}}},\ }\bibfield
  {title} {\bibinfo {title} {The resource theory of informational
  nonequilibrium in thermodynamics},\ }\href
  {https://doi.org/10.1016/j.physrep.2015.04.003} {\bibfield  {journal}
  {\bibinfo  {journal} {Physics Reports}\ }\textbf {\bibinfo {volume} {583}},\
  \bibinfo {pages} {1} (\bibinfo {year} {2015})}\BibitemShut {NoStop}%
\bibitem [{\citenamefont {Horodecki}\ and\ \citenamefont
  {Oppenheim}(2013{\natexlab{b}})}]{HORODECKIdoi:10.1142/S0217979213450197}%
  \BibitemOpen
  \bibfield  {author} {\bibinfo {author} {\bibfnamefont {M.}~\bibnamefont
  {Horodecki}}\ and\ \bibinfo {author} {\bibfnamefont {J.}~\bibnamefont
  {Oppenheim}},\ }\bibfield  {title} {\bibinfo {title} {{(Quantumness in the
  context of) Resource Theories}},\ }\href
  {https://doi.org/10.1142/S0217979213450197} {\bibfield  {journal} {\bibinfo
  {journal} {Int. J. Mod. Phys. B}\ }\textbf {\bibinfo {volume} {27}},\
  \bibinfo {pages} {1345019} (\bibinfo {year}
  {2013}{\natexlab{b}})}\BibitemShut {NoStop}%
\bibitem [{\citenamefont {Bennett}\ \emph
  {et~al.}(1996{\natexlab{c}})\citenamefont {Bennett}, \citenamefont
  {DiVincenzo}, \citenamefont {Smolin},\ and\ \citenamefont
  {Wootters}}]{BennettPhysRevA.54.3824}%
  \BibitemOpen
  \bibfield  {author} {\bibinfo {author} {\bibfnamefont {C.~H.}\ \bibnamefont
  {Bennett}}, \bibinfo {author} {\bibfnamefont {D.~P.}\ \bibnamefont
  {DiVincenzo}}, \bibinfo {author} {\bibfnamefont {J.~A.}\ \bibnamefont
  {Smolin}},\ and\ \bibinfo {author} {\bibfnamefont {W.~K.}\ \bibnamefont
  {Wootters}},\ }\bibfield  {title} {\bibinfo {title} {Mixed-state entanglement
  and quantum error correction},\ }\href
  {https://doi.org/10.1103/PhysRevA.54.3824} {\bibfield  {journal} {\bibinfo
  {journal} {Phys. Rev. A}\ }\textbf {\bibinfo {volume} {54}},\ \bibinfo
  {pages} {3824} (\bibinfo {year} {1996}{\natexlab{c}})}\BibitemShut {NoStop}%
\bibitem [{\citenamefont {Werner}(1989)}]{WernerPhysRevA.40.4277}%
  \BibitemOpen
  \bibfield  {author} {\bibinfo {author} {\bibfnamefont {R.~F.}\ \bibnamefont
  {Werner}},\ }\bibfield  {title} {\bibinfo {title} {{Quantum states with
  Einstein-Podolsky-Rosen correlations admitting a hidden-variable model}},\
  }\href {https://doi.org/10.1103/PhysRevA.40.4277} {\bibfield  {journal}
  {\bibinfo  {journal} {Phys. Rev. A}\ }\textbf {\bibinfo {volume} {40}},\
  \bibinfo {pages} {4277} (\bibinfo {year} {1989})}\BibitemShut {NoStop}%
\bibitem [{\citenamefont {Baumgratz}\ \emph {et~al.}(2014)\citenamefont
  {Baumgratz}, \citenamefont {Cramer},\ and\ \citenamefont
  {Plenio}}]{BaumgratzPhysRevLett.113.140401}%
  \BibitemOpen
  \bibfield  {author} {\bibinfo {author} {\bibfnamefont {T.}~\bibnamefont
  {Baumgratz}}, \bibinfo {author} {\bibfnamefont {M.}~\bibnamefont {Cramer}},\
  and\ \bibinfo {author} {\bibfnamefont {M.~B.}\ \bibnamefont {Plenio}},\
  }\bibfield  {title} {\bibinfo {title} {{Quantifying Coherence}},\ }\href
  {https://doi.org/10.1103/PhysRevLett.113.140401} {\bibfield  {journal}
  {\bibinfo  {journal} {Phys. Rev. Lett.}\ }\textbf {\bibinfo {volume} {113}},\
  \bibinfo {pages} {140401} (\bibinfo {year} {2014})}\BibitemShut {NoStop}%
\bibitem [{Note1()}]{Note1}%
  \BibitemOpen
  \bibinfo {note} {Throughout this article, we set $k=1$.}\BibitemShut {Stop}%
\bibitem [{\citenamefont {Vidal}\ and\ \citenamefont
  {Cirac}(2001)}]{VidalPhysRevLett.86.5803}%
  \BibitemOpen
  \bibfield  {author} {\bibinfo {author} {\bibfnamefont {G.}~\bibnamefont
  {Vidal}}\ and\ \bibinfo {author} {\bibfnamefont {J.~I.}\ \bibnamefont
  {Cirac}},\ }\bibfield  {title} {\bibinfo {title} {{Irreversibility in
  Asymptotic Manipulations of Entanglement}},\ }\href
  {https://doi.org/10.1103/PhysRevLett.86.5803} {\bibfield  {journal} {\bibinfo
   {journal} {Phys. Rev. Lett.}\ }\textbf {\bibinfo {volume} {86}},\ \bibinfo
  {pages} {5803} (\bibinfo {year} {2001})}\BibitemShut {NoStop}%
\bibitem [{\citenamefont {Lami}\ and\ \citenamefont
  {Regula}(2021)}]{Lami2111.02438}%
  \BibitemOpen
  \bibfield  {author} {\bibinfo {author} {\bibfnamefont {L.}~\bibnamefont
  {Lami}}\ and\ \bibinfo {author} {\bibfnamefont {B.}~\bibnamefont {Regula}},\
  }\bibfield  {title} {\bibinfo {title} {No second law of entanglement
  manipulation after all},\ }\href {https://arxiv.org/abs/2111.02438}
  {\bibfield  {journal} {\bibinfo  {journal} {arXiv:2111.02438}\ } (\bibinfo
  {year} {2021})}\BibitemShut {NoStop}%
\bibitem [{\citenamefont {Horodecki}\ \emph {et~al.}(2000)\citenamefont
  {Horodecki}, \citenamefont {Horodecki},\ and\ \citenamefont
  {Horodecki}}]{HorodeckiPhysRevLett.84.2014}%
  \BibitemOpen
  \bibfield  {author} {\bibinfo {author} {\bibfnamefont {M.}~\bibnamefont
  {Horodecki}}, \bibinfo {author} {\bibfnamefont {P.}~\bibnamefont
  {Horodecki}},\ and\ \bibinfo {author} {\bibfnamefont {R.}~\bibnamefont
  {Horodecki}},\ }\bibfield  {title} {\bibinfo {title} {{Limits for
  Entanglement Measures}},\ }\href
  {https://doi.org/10.1103/PhysRevLett.84.2014} {\bibfield  {journal} {\bibinfo
   {journal} {Phys. Rev. Lett.}\ }\textbf {\bibinfo {volume} {84}},\ \bibinfo
  {pages} {2014} (\bibinfo {year} {2000})}\BibitemShut {NoStop}%
\bibitem [{\citenamefont {Devetak}\ and\ \citenamefont
  {Winter}(2005)}]{Devetak2005}%
  \BibitemOpen
  \bibfield  {author} {\bibinfo {author} {\bibfnamefont {I.}~\bibnamefont
  {Devetak}}\ and\ \bibinfo {author} {\bibfnamefont {A.}~\bibnamefont
  {Winter}},\ }\bibfield  {title} {\bibinfo {title} {Distillation of secret key
  and entanglement from quantum states},\ }\href
  {https://doi.org/10.1098/rspa.2004.1372} {\bibfield  {journal} {\bibinfo
  {journal} {Proceedings of the Royal Society A: Mathematical, Physical and
  Engineering Sciences}\ }\textbf {\bibinfo {volume} {461}},\ \bibinfo {pages}
  {207} (\bibinfo {year} {2005})}\BibitemShut {NoStop}%
\bibitem [{\citenamefont {Pal}\ \emph {et~al.}(2014)\citenamefont {Pal},
  \citenamefont {Bandyopadhyay},\ and\ \citenamefont
  {Ghosh}}]{PalPhysRevA.90.052304}%
  \BibitemOpen
  \bibfield  {author} {\bibinfo {author} {\bibfnamefont {R.}~\bibnamefont
  {Pal}}, \bibinfo {author} {\bibfnamefont {S.}~\bibnamefont {Bandyopadhyay}},\
  and\ \bibinfo {author} {\bibfnamefont {S.}~\bibnamefont {Ghosh}},\ }\bibfield
   {title} {\bibinfo {title} {Entanglement sharing through noisy qubit
  channels: One-shot optimal singlet fraction},\ }\href
  {https://doi.org/10.1103/PhysRevA.90.052304} {\bibfield  {journal} {\bibinfo
  {journal} {Phys. Rev. A}\ }\textbf {\bibinfo {volume} {90}},\ \bibinfo
  {pages} {052304} (\bibinfo {year} {2014})}\BibitemShut {NoStop}%
\bibitem [{\citenamefont {Streltsov}\ \emph {et~al.}(2015)\citenamefont
  {Streltsov}, \citenamefont {Augusiak}, \citenamefont {Demianowicz},\ and\
  \citenamefont {Lewenstein}}]{StreltsovPhysRevA.92.012335}%
  \BibitemOpen
  \bibfield  {author} {\bibinfo {author} {\bibfnamefont {A.}~\bibnamefont
  {Streltsov}}, \bibinfo {author} {\bibfnamefont {R.}~\bibnamefont {Augusiak}},
  \bibinfo {author} {\bibfnamefont {M.}~\bibnamefont {Demianowicz}},\ and\
  \bibinfo {author} {\bibfnamefont {M.}~\bibnamefont {Lewenstein}},\ }\bibfield
   {title} {\bibinfo {title} {Progress towards a unified approach to
  entanglement distribution},\ }\href
  {https://doi.org/10.1103/PhysRevA.92.012335} {\bibfield  {journal} {\bibinfo
  {journal} {Phys. Rev. A}\ }\textbf {\bibinfo {volume} {92}},\ \bibinfo
  {pages} {012335} (\bibinfo {year} {2015})}\BibitemShut {NoStop}%
\bibitem [{\citenamefont {King}(2002)}]{King2022}%
  \BibitemOpen
  \bibfield  {author} {\bibinfo {author} {\bibfnamefont {C.}~\bibnamefont
  {King}},\ }\bibfield  {title} {\bibinfo {title} {Additivity for unital qubit
  channels},\ }\href {https://doi.org/10.1063/1.1500791} {\bibfield  {journal}
  {\bibinfo  {journal} {Journal of Mathematical Physics}\ }\textbf {\bibinfo
  {volume} {43}},\ \bibinfo {pages} {4641} (\bibinfo {year}
  {2002})}\BibitemShut {NoStop}%
\bibitem [{\citenamefont {Nielsen}\ and\ \citenamefont
  {Chuang}(2010)}]{nielsen2002quantum}%
  \BibitemOpen
  \bibfield  {author} {\bibinfo {author} {\bibfnamefont {M.~A.}\ \bibnamefont
  {Nielsen}}\ and\ \bibinfo {author} {\bibfnamefont {I.~L.}\ \bibnamefont
  {Chuang}},\ }\href {https://doi.org/10.1017/CBO9780511976667} {\emph
  {\bibinfo {title} {Quantum Computation and Quantum Information: 10th
  Anniversary Edition}}}\ (\bibinfo  {publisher} {Cambridge University Press},\
  \bibinfo {year} {2010})\BibitemShut {NoStop}%
\bibitem [{\citenamefont {Wootton}\ and\ \citenamefont
  {Loss}(2018)}]{wootton2018repetition}%
  \BibitemOpen
  \bibfield  {author} {\bibinfo {author} {\bibfnamefont {J.~R.}\ \bibnamefont
  {Wootton}}\ and\ \bibinfo {author} {\bibfnamefont {D.}~\bibnamefont {Loss}},\
  }\bibfield  {title} {\bibinfo {title} {Repetition code of 15 qubits},\ }\href
  {https://doi.org/10.1103/PhysRevA.97.052313} {\bibfield  {journal} {\bibinfo
  {journal} {Phys. Rev. A}\ }\textbf {\bibinfo {volume} {97}},\ \bibinfo
  {pages} {052313} (\bibinfo {year} {2018})}\BibitemShut {NoStop}%
\bibitem [{\citenamefont {Simakov}\ \emph {et~al.}(2022)\citenamefont
  {Simakov}, \citenamefont {Besedin},\ and\ \citenamefont
  {Ustinov}}]{simakov2022simulation}%
  \BibitemOpen
  \bibfield  {author} {\bibinfo {author} {\bibfnamefont {I.~A.}\ \bibnamefont
  {Simakov}}, \bibinfo {author} {\bibfnamefont {I.~S.}\ \bibnamefont
  {Besedin}},\ and\ \bibinfo {author} {\bibfnamefont {A.~V.}\ \bibnamefont
  {Ustinov}},\ }\bibfield  {title} {\bibinfo {title} {Simulation of the
  five-qubit quantum error correction code on superconducting qubits},\ }\href
  {https://doi.org/10.1103/PhysRevA.105.032409} {\bibfield  {journal} {\bibinfo
   {journal} {Phys. Rev. A}\ }\textbf {\bibinfo {volume} {105}},\ \bibinfo
  {pages} {032409} (\bibinfo {year} {2022})}\BibitemShut {NoStop}%
\end{thebibliography}%

\section*{Supplemental Material}
\subsection{Asymptotic state conversion \protect \\
 in general quantum resource theories}

A general quantum resource theory is defined via a set of free states
$\mathcal{F}$ and the set of free operations $\{\Lambda_{f}\}$.
Any state which is not element of $\mathcal{F}$ is called resource
state. Typically, $\mathcal{F}$ is a convex subset of all quantum
states. Moreover, the free operations $\Lambda_{f}$ cannot convert
free states into resource states. For a general resource theory, asymptotic
conversion rate between two resource states $\rho$ and $\sigma$
can be defined as~\cite{HORODECKIdoi:10.1142/S0217979213450197}
\begin{equation}
R(\rho\rightarrow\sigma)=\sup\left\{ r:\lim_{n\rightarrow\infty}\left(\inf_{\left\{ \Lambda_{f}\right\} }\left\Vert \Lambda_{f}\left[\rho^{\otimes n}\right]-\sigma^{\otimes\left\lfloor rn\right\rfloor }\right\Vert _{1}\right)=0\right\} 
\end{equation}
with the trace norm $||M||_{1}=\mathrm{Tr}\sqrt{M^{\dagger}M}$.

In the resource theory of entanglement, free operations are typically
assumed to be local operations and classical communication (LOCC),
corresponding to a setting where the remote parties can apply general
local measurements, and communicate the outcomes of the measurements
via a classical channel~\cite{BennettPhysRevA.54.3824}. The free
states of this theory are separable states $\rho_{\mathrm{sep}}=\sum_{i}p_{i}\rho_{i}^{A}\otimes\rho_{i}^{B}$~\cite{WernerPhysRevA.40.4277}. In this theory, asymptotic conversion rates are known
for transformations between bipartite pure states~\cite{BennettPhysRevA.53.2046}:
\begin{equation}
R(\psi^{AB}\rightarrow\phi^{AB})=\frac{S(\psi^{A})}{S(\phi^{A})}.
\end{equation}
The general conversion rates are related to the distillable entanglement and entanglement cost~\cite{Plenioquant-ph/0504163} of a state $\rho$ as follows:
\begin{align}
E_{\mathrm{d}}(\rho) & =R(\rho\rightarrow\psi^{-})\\
E_{\mathrm{c}}(\rho) & =\frac{1}{R(\psi^{-}\rightarrow\rho)}
\end{align}

In the resource theory of coherence, 
the free states are diagonal
in a reference basis
$\{\ket{i}\}$~\cite{Abergquant-ph/0612146,BaumgratzPhysRevLett.113.140401,StreltsovRevModPhys.89.041003}.
The set of all states diagonal in the reference basis will be denoted
by $\mathcal{I}$.
While different sets of free operations have been
defined in the
literature~\cite{StreltsovRevModPhys.89.041003},
we focus on maximally incoherent operations in this article.
These are all operations which cannot create coherence from diagonal states,
i.e., $\Lambda[\sigma]\in\mathcal{I}$ for any $\sigma\in\mathcal{I}$~\cite{Abergquant-ph/0612146}.
Asymptotic conversion rates under maximally incoherent operations
are known for transition between all quantum states~\cite{WinterPhysRevLett.116.120404}:
\begin{equation}
R(\rho\rightarrow\sigma)=\frac{C(\rho)}{C(\sigma)}
\end{equation}
with the relative entropy of coherence~\cite{BaumgratzPhysRevLett.113.140401}
\begin{equation}
C(\rho)=\inf_{\sigma\in\mathcal{I}}S(\rho||\sigma)=S(\Delta[\rho])-S(\rho).
\end{equation}

In the resource theory of quantum thermodynamics, we consider a quantum
system $S$ with Hamiltonian $H_{S}$. The free state is the Gibbs
state 
\begin{equation}
\gamma^{S}=\frac{e^{-\beta H_{S}}}{\mathrm{Tr}[e^{-\beta H_{S}}]}
\end{equation}
with the inverse temperature $\beta=1/kT$ \footnote{Throughout this article, we set $k=1$.}.
The free operations of this theory can be implemented by coupling
the system to a thermal bath with Hamiltonian $H_{B}$ and applying
an energy preserving unitary~\cite{Janzing2000}: 
\begin{equation}
\Lambda[\rho^{S}]=\mathrm{Tr}_{B}[U(\rho^{S}\otimes\gamma^{B})U^{\dagger}],\label{eq:ThermalOperations}
\end{equation}
where $\gamma^{B}$ is the Gibbs state of the bath, and the unitary
$U$ fulfills $[U,H_{S}+H_{B}]=0$. Transformations of the form~(\ref{eq:ThermalOperations})
are known as thermal operations. Asymptotic conversion rates in this
theory are given by~\cite{BrandaoPhysRevLett.111.250404} 
\begin{equation}
R(\rho\rightarrow\sigma)=\frac{S(\rho||\gamma)}{S(\sigma||\gamma)},
\end{equation}
where the final state $\sigma$ is diagonal in the eigenbasis of the
Hamiltonian. For a general state $\sigma$ conversion is possible
at the same rate if a sublinear number of qubits with coherence is
provided~\cite{BrandaoPhysRevLett.111.250404}.

A resource theory is called reversible if asymptotic transitions between
any resource states are possible in a lossless way, i.e., if for any
pair of resource states $\rho$ and $\sigma$ the following equality
holds \cite{HORODECKIdoi:10.1142/S0217979213450197} 
\begin{equation}
R(\rho\rightarrow\sigma)\times R(\sigma\rightarrow\rho)=1. \label{eq:Reversibility}
\end{equation}
From the above discussion it is clear that the resource theory of
purity and the resource theory of coherence based on maximally incoherent
operations are reversible. While the resource theory of entanglement
is reversible for bipartite pure states, it is not reversible in general~\cite{VidalPhysRevLett.86.5803},
even if all non-entangling operations are taken into account~\cite{Lami2111.02438}. In the following, we assume that asymptotic transformations between any two resource states are possible with non-zero rate, as is the case in most quantum resource theories.

Note that for any reversible theory, the following equality holds for any three resource states $\rho$, $\sigma$, and $\tau$:
\begin{equation}
R(\rho\rightarrow\sigma)\times R(\sigma\rightarrow\tau)=R(\rho\rightarrow\tau). \label{eq:Reversibility3States}
\end{equation}
This can be seen by contradiction, assuming that the equality does not hold for some resource states. Let us first assume that 
\begin{equation}
R(\rho\rightarrow\sigma)\times R(\sigma\rightarrow\tau) < R(\rho\rightarrow\tau),
\end{equation}
which implies that 
\begin{equation}
R(\rho\rightarrow\sigma)<\frac{R(\rho\rightarrow\tau)}{R(\sigma\rightarrow\tau)}=R(\rho\rightarrow\tau)\times R(\tau\rightarrow\sigma),
\end{equation}
where the last equality was obtained by using Eq.~(\ref{eq:Reversibility}). This is a contradiction, since it is always possible to convert $\rho$ into $\sigma$ via the state $\tau$, i.e., $R(\rho\rightarrow\sigma)\geq R(\rho\rightarrow\tau)\times R(\tau\rightarrow\sigma)$. The remaining case
\begin{equation}
R(\rho\rightarrow\sigma)\times R(\sigma\rightarrow\tau) > R(\rho\rightarrow\tau)
\end{equation}
is treated similarly. Also in this case we obtain a contradiction, as it is always possible to convert $\rho$ into $\tau$ via the state $\sigma$.

Using the above results, we can show that for any reversible resource theory the performance
of the dilution procedure does not depend on the initial state, i.e.,
\begin{equation}
R\left(\rho\rightarrow\mu\right)\times R\left(\Lambda\left[\mu\right]\rightarrow\rho\right)=R\left(\sigma\rightarrow\mu\right)\times R\left(\Lambda\left[\mu\right]\rightarrow\sigma\right)
\end{equation}
for any two resource states $\rho$ and $\sigma$. This can be seen directly using Eq.~(\ref{eq:Reversibility3States}):
\begin{align}
R\left(\rho\rightarrow\mu\right)\times R\left(\Lambda\left[\mu\right]\rightarrow\rho\right) & =R\left(\rho\rightarrow\sigma\right)\times R\left(\sigma\rightarrow\mu\right)\nonumber \\
 & \times R\left(\Lambda\left[\mu\right]\rightarrow\sigma\right)\times R\left(\sigma\rightarrow\rho\right) \nonumber\\
 & =R\left(\sigma\rightarrow\mu\right)\times R\left(\Lambda\left[\mu\right]\rightarrow\sigma\right).
\end{align}

Under few minimal assumptions, it was shown in~\cite{HORODECKIdoi:10.1142/S0217979213450197}
that for any reversible resource theory the asymptotic conversion
rates are given by 
\begin{equation}
R(\rho\rightarrow\sigma)=\frac{E_{\mathrm r}^{\infty}(\rho)}{E_{\mathrm r}^{\infty}(\sigma)},\label{eq:ReversibleRate}
\end{equation}
where $E_{\mathrm r}^{\infty}(\rho)=\lim_{n\rightarrow\infty}E_{\mathrm r}(\rho^{\otimes n})/n$,
and $E_{\mathrm r}(\rho)$ denotes the minimal relative entropy between $\rho$
and the set of free states $\mathcal{F}$: $E_{\mathrm r}(\rho)=\inf_{\sigma\in\mathcal{F}}S(\rho||\sigma)$. This applies to the resource theory of purity and the resource theory of coherence based on maximally incoherent operations. Also, in the resource theory of quantum thermodynamics the conversion rates are given by expressions of the form (\ref{eq:ReversibleRate}).

\subsection{Advantage of entanglement dilution}

\subsubsection{Single-qubit noise}
We consider single-qubit phase damping on Bob's side with Kraus operators
as given in Eq.~(2) of the main text. If no dilution is applied,
after the action of the noise Alice and Bob can obtain singlets at
rate $E_{\mathrm{d}}(\openone\otimes\Lambda[\psi^{-}])$. Since the
state $\openone\otimes\Lambda[\psi^{-}]$ is maximally correlated,
we can evaluate its distillable entanglement~\cite{HorodeckiPhysRevLett.84.2014,Devetak2005}:
\begin{equation}
E_{\mathrm{d}}\left(\openone\otimes\Lambda[\psi^{-}]\right)=1-h\left(\frac{1+\sqrt{1-\lambda}}{2}\right)
\end{equation}
with the binary entropy $h(x)=-x\log_{2}x-(1-x)\log_{2}(1-x)$.

We will now show that a better performance can be achieved if Alice
and Bob first dilute their singlets into states of the form $\ket{\psi}=\cos\alpha\ket{00}+\sin\alpha\ket{11}$.
The state $\openone\otimes\Lambda[\psi]$ is also maximally correlated,
and its distillable entanglement is given as~\cite{HorodeckiPhysRevLett.84.2014,Devetak2005}
\begin{equation}
E_{\mathrm{d}}\left(\openone\otimes\Lambda\left[\psi\right]\right)=h\left(\cos^{2}\alpha\right)-h\left(\frac{1}{2}+\frac{\sqrt{2\lambda\cos(4\alpha)-2\lambda+4}}{4}\right).
\label{eq:distillable_entanglement_advantagedilution}
\end{equation}

With these results, we can evaluate both sides of the inequality~(1) of the main text,
the result is shown in Fig.~2 of the main text.

\subsubsection{Noise of larger dimension}
\label{sec:2QubitNoise}

We will now consider two-qubit noise of the form 
\begin{align}
\Lambda[\rho] & =p_{00}\rho+p_{01}\left(\openone\otimes\sigma_{z}\right)\rho\left(\openone\otimes\sigma_{z}\right) \label{eq:2QubitNoise-1} \\
 & +p_{10}\left(\sigma_{z}\otimes\openone\right)\rho\left(\sigma_{z}\otimes\openone\right)+p_{11}\left(\sigma_{z}\otimes\sigma_{z}\right)\rho\left(\sigma_{z}\otimes\sigma_{z}\right)\nonumber 
\end{align}
with $\sum_{i,j}p_{ij}=1$. Assume now that Alice and Bob share $2$
singlets and the noise acts only on Bob's part. The state after the
action of the noise can be written as 
\begin{align}
\Lambda_{B}[(\psi^{-})^{\otimes2}] & =p_{00}\psi^{-}\otimes\psi^{-}+p_{01}\psi^{-}\otimes\psi^{+}\\
 & +p_{10}\psi^{+}\otimes\psi^{-}+p_{11}\psi^{+}\otimes\psi^{+} \nonumber 
\end{align}
with $\ket{\psi^+} = (\ket{01}+\ket{10})/\sqrt{2}$. Noting that this state is maximally correlated, the distillable entanglement
of the state can be evaluated as follows:
\begin{equation}
E_{\mathrm{d}}(\Lambda_B[(\psi^{-})^{\otimes2}])=2-h(\boldsymbol{p})
\end{equation}
with the probability vector $\boldsymbol{p}=(p_{00},p_{01},p_{10},p_{11})$, and the Shannon entropy $h(\boldsymbol{p})=-\sum_{i,j} p_{ij}\log_{2}p_{ij}$.

Assume now that Alice and Bob dilute their singlets into states of
the form $\ket{\psi}=\cos\alpha\ket{00}+\sin\alpha\ket{11}$ before
the action of the noise. In the asymptotic limit, the dilution provides
an advantage whenever the inequality 
\begin{equation}
\frac{E_{\mathrm{d}}(\Lambda_{B}[\psi^{\otimes2}])}{S(\psi^{A})}>E_{\mathrm{d}}(\Lambda_B[(\psi^{-})^{\otimes2}]) \label{eq:2QubitNoise}
\end{equation}
is fulfilled. Since the state $\Lambda_{B}[\psi^{\otimes2}]$
is maximally correlated, we can evaluate the distillable entanglement
as follows:
\begin{equation}
E_{\mathrm{d}}(\Lambda_{B}[\psi^{\otimes2}])=2h(\cos^{2}\alpha)-S(\Lambda_{B}[\psi^{\otimes2}]).
\end{equation}

\begin{figure}
\includegraphics[width=1\columnwidth]{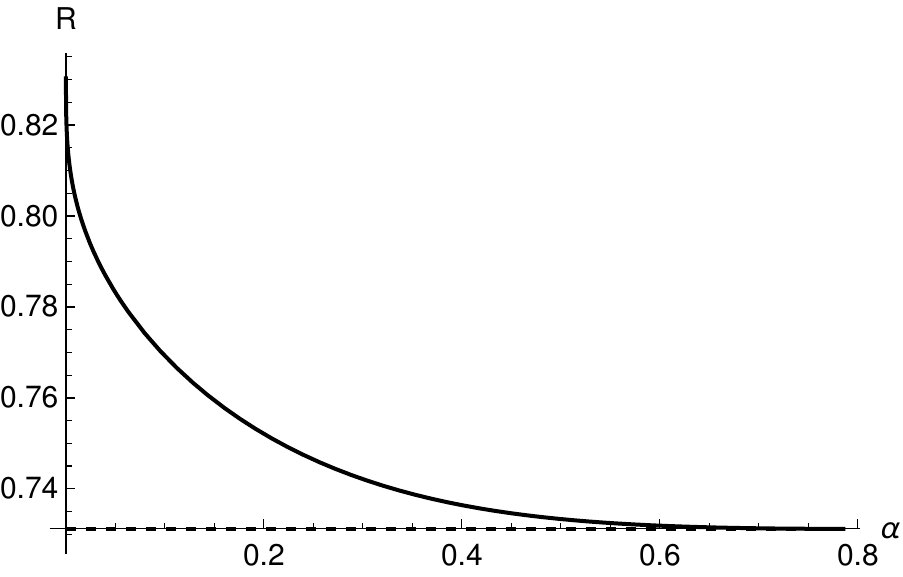}

\caption{Reducing entanglement loss under local two-qubit noise by diluting singlets into pure states $\ket{\psi} = \cos \alpha \ket{00} + \sin \alpha \ket{11}$. Solid curve shows $E_{\mathrm{d}}(\Lambda_{B}[\psi^{\otimes2}])/S(\psi^A)$ as a function of $\alpha$, where the noise parameters are given in Section~\ref{sec:2QubitNoise}. Dashed line shows $E_{\mathrm{d}}(\Lambda_B[(\psi^{-})^{\otimes2}])$.}
\label{fig:2QubitNoise} 
\end{figure}

We have performed a numerical comparison of both sides of the inequality~(\ref{eq:2QubitNoise}) for different probabilities $p_i$ and parameters $\alpha$, finding that in many cases dilution into a weakly entangled state provides an advantage. As an example, we show in Fig.~\ref{fig:2QubitNoise} both sides of the inequality~(\ref{eq:2QubitNoise}) as a function of $\alpha$ for $p_{00}=0.05$, $p_{01}=0.03$, $p_{10}=0.26$, $p_{11}=0.66$. Similar to the single-qubit noise, optimal performance is found in the limit of a non-entangled state $\alpha \rightarrow 0$. 

\begin{figure}
\includegraphics[width=1\columnwidth]{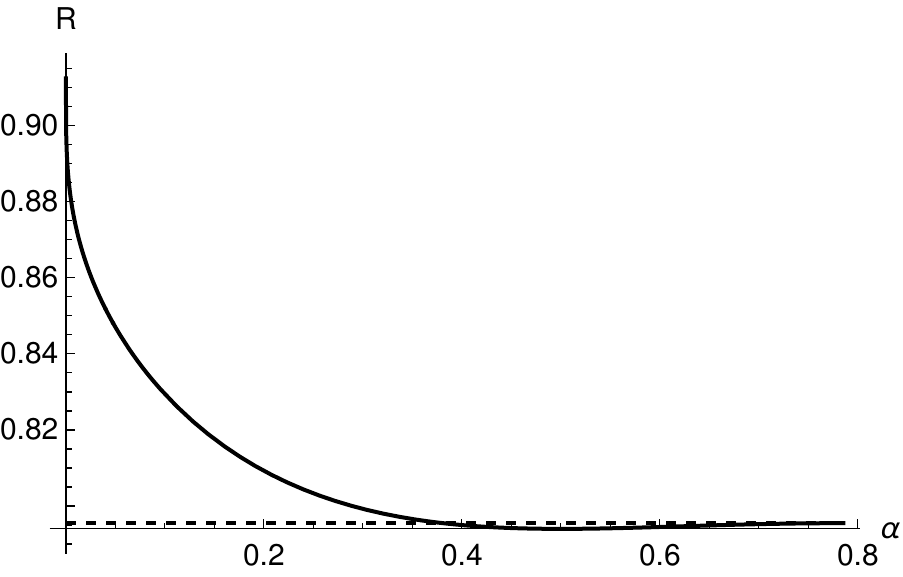}

\caption{Reducing entanglement loss under local three-qubit noise by diluting singlets into pure states $\ket{\psi} = \cos \alpha \ket{00} + \sin \alpha \ket{11}$. Solid curve shows $E_{\mathrm{d}}(\Lambda_{B}[\psi^{\otimes 3}])/S(\psi^A)$ as a function of $\alpha$. Dashed line shows $E_{\mathrm{d}}(\Lambda_B[(\psi^{-})^{\otimes 3}])$.}
\label{fig:3QubitNoise} 
\end{figure}

The analysis presented above can be extended to $n$ qubits by extending the noise in Eq.~(\ref{eq:2QubitNoise-1}) in a straightforward way, e.g. for $n=3$ the noise can be written as 
\begin{equation}
\Lambda[\rho]=\sum_{k,l,m\in\{0,1\}}p_{klm}\left(\sigma_{z}^{k}\otimes\sigma_{z}^{l}\otimes\sigma_{z}^{m}\right)\rho\left(\sigma_{z}^{k}\otimes\sigma_{z}^{l}\otimes\sigma_{z}^{m}\right)
\end{equation}
with $\sum_{k,l,m\in\{0,1\}}p_{klm}=1$. Dilution into states $\ket{\psi} = \cos \alpha \ket{00} + \sin \alpha \ket{11}$ provides an advantage whenever
\begin{equation}
\frac{E_{\mathrm{d}}(\Lambda_{B}[\psi^{\otimes n}])}{S(\psi^{A})}>E_{\mathrm{d}}(\Lambda_B[(\psi^{-})^{\otimes n}]). \label{eq:nQubits}
\end{equation}
Noting that the states $\Lambda_{B}[(\psi^{-})^{\otimes n}]$ and $\Lambda_{B}[\psi^{\otimes n}]$ are maximally correlated, we further obtain 
\begin{align}
E_{\mathrm{d}}(\Lambda_{B}[(\psi^{-})^{\otimes n}]) & =n-h(\boldsymbol{p}),\\
E_{\mathrm{d}}(\Lambda_{B}[\psi^{\otimes n}]) & =n h(\cos^{2}\alpha)-S(\Lambda_{B}[\psi^{\otimes n}]).
\end{align}
In Fig.~\ref{fig:3QubitNoise} we show both sides of the inequality~(\ref{eq:nQubits}) as a function of $\alpha$ for $n=3$ with $p_{000}=0.06$, $p_{001}=0.03$, $p_{010}=0.04$, $p_{100}=0.31$, $p_{011}=0.01$, $p_{101}=0.42$, $p_{110}=0.05$, and $p_{111}=0.08$. Also in this case we find that optimal performance is achieved in the limit $\alpha \rightarrow 0$. We have also performed numerics for $n=4$, finding that also in this case there is a parameter range for the noise where dilution provides an advantage.

\subsubsection{Comparison to results on entanglement distribution}

It is instrumental to compare the findings presented above
to earlier results investigating the role of weakly entangled states
for entanglement distribution. As has been noticed in~\cite{PalPhysRevA.90.052304,StreltsovPhysRevA.92.012335},
for some noisy qubit channels $\Lambda$ there exist non-maximally
entangled states $\ket{\psi}$ such that $E(\openone\otimes\Lambda[\psi])>E(\openone\otimes\Lambda[\psi^{-}])$,
where $E$ is a suitably chosen quantifier of entanglement. As has
been further shown in~\cite{StreltsovPhysRevA.92.012335}, this effect
never appears for Pauli channels, i.e., a quantum channels of the
type $\Lambda[\rho]=\sum_{i=0}^{3}p_{i}\sigma_{i}\rho\sigma_{i}$
with $\sigma_{0}=\openone$. In this situation, it is true that $E(\openone\otimes\Lambda[\psi^{-}])\geq E(\openone\otimes\Lambda[\rho])$
for any bipartite state $\rho$ and any entanglement quantifier $E$~\cite{StreltsovPhysRevA.92.012335}.
Noting that the noise defined in Eq.~(2) of the main text also
corresponds to a Pauli channel, we see that the problem considered
in our article is significantly different from entanglement distribution
via noisy channels. While maximally entangled states are optimal resource
for entanglement distribution via a noisy channel of the form~(2) of the main text,
they are not necessarily an optimal choice to reduce the entanglement
loss under the same type of noise. 

\subsection{Minimal loss of coherence for pure states and single-qubit amplitude damping noise}

In Fig.~3 of the main text we show the overall coherence rate $C(\Lambda[\psi])/C(\psi)$ for single-qubit amplitude damping noise and pure states of the form 
\begin{equation}
    \ket{\psi} = \cos \alpha \ket{0} + \sin \alpha \ket{1} \label{eq:CoherencePureProof}
\end{equation}
as a function of $\alpha$, where the maximal rate $R \approx 0.15$ was attained for $\alpha \approx 0.34$. We will now show that this is indeed the maximal possible rate achievable by coherence dilution into pure qubit states. For this, it is enough to notice that any other qubit state can be obtained from a state of the form~(\ref{eq:CoherencePureProof}) by applying a diagonal unitary. Note that the amplitude damping noise commutes with any such unitary, and moreover the relative entropy of coherence $C$ is invariant under digonal unitaries. We thus obtain 
\begin{equation}
\frac{C(\Lambda[U\psi U^{\dagger}])}{C(U\psi U^{\dagger})}=\frac{C(\Lambda[\psi])}{C(\psi)},
\end{equation}
which proves the claim.

\subsection{Quantum thermodynamics}

\subsubsection{Optimal states for thermal noise in Eq.~(6) of the main text}

\begin{figure}
\includegraphics[width=1\columnwidth]{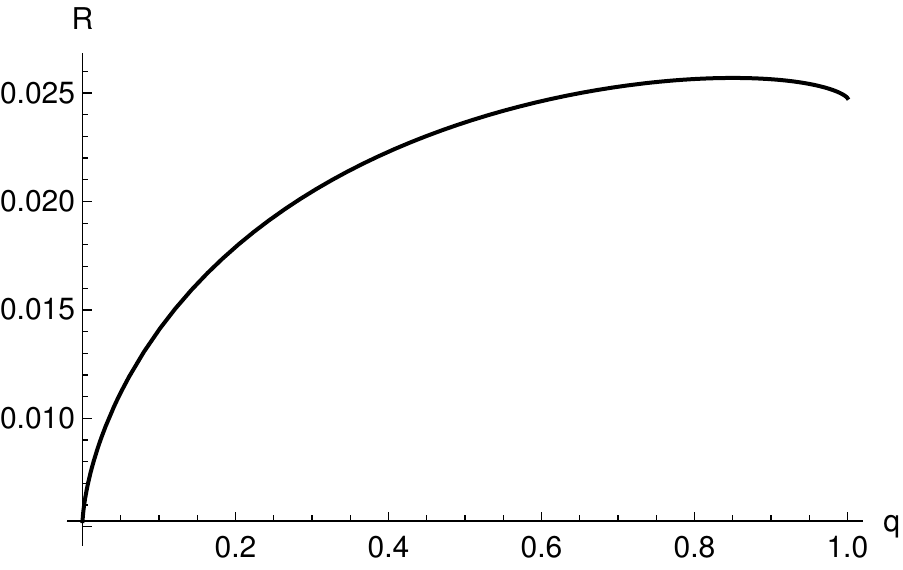}

\caption{Reducing the loss of free energy of the excited states under thermal
noise given in Eq.~(6) of the main text by diluting them into
diagonal states of the form~(7) of the main text. The curve
shows the final rate of excited states $R=S\left(\Lambda\left[\mu\right]||\gamma\right)/S\left(\mu||\gamma\right)$
as a function of $q$ for $T=0.3$ and $p=0.9$. Maximal performance
is achieved for $q\approx0.85$.}
\label{fig:Thermal} 
\end{figure}

Here we will prove that for noise of the form~(6) of the main text
the optimal strategy is to dilute the excited state into a diagonal
state of the form~(7) of the main text. For proving this, note
that $\Lambda$ has the following property: $\Lambda(\Delta[\rho])=\Lambda[\rho]$.
This implies that 
\begin{equation}
S(\Lambda(\Delta[\mu])||\gamma)=S(\Lambda[\mu]||\gamma).
\end{equation}
for any state $\mu$. In the next step, we express the quantum relative
entropy as follows: 
\begin{align}
S(\rho||\gamma)=-S(\rho)-\mathrm{Tr}\left[\Delta\left(\rho\right)\log_{2}\gamma\right].
\end{align}
Recall that the von Neumann entropy does not decrease under dephasing:
$S(\Delta[\mu])\geq S(\mu)$, with equality if and only if $\Delta[\mu]=\mu$.
It follows that 
\begin{align}
S\left(\Delta\left[\mu\right]||\gamma\right) & =-S\left(\Delta\left[\mu\right]\right)-\mathrm{Tr}\left[\Delta\left(\mu\right)\log_{2}\gamma\right]\\
 & \leq-S\left(\mu\right)-\mathrm{Tr}\left[\Delta\left(\mu\right)\log_{2}\gamma\right]=S\left(\mu||\gamma\right)\nonumber 
\end{align}
with equality if and only if $\Delta[\mu]=\mu$. Combining the above
results, we see that 
\begin{equation}
\frac{S\left(\Lambda\left[\mu\right]||\gamma\right)}{S\left(\mu||\gamma\right)}\leq\frac{S\left(\Lambda\left[\Delta\left[\mu\right]\right]||\gamma\right)}{S\left(\Delta\left[\mu\right]||\gamma\right)}.
\end{equation}
For $p<1$ equality is attained if and only if $\mu$ is diagonal
in the eigenbasis of the Hamiltonian. This proves that optimal performance
is achieved on diagonal states of the form~(7).

Our results for the resource theories of entanglement and
coherence suggest that dilution procedure shows optimal performance
in the limit of the resource-free states. We will see that this is
no longer the case in quantum thermodynamics, where the optimal states
can be far away from the Gibbs state. This can be seen from Fig.~\ref{fig:Thermal},
where we show the overall rate $S\left(\Lambda\left[\mu\right]||\gamma\right)/S\left(\mu||\gamma\right)$
for diagonal states $\mu$ as a function of $q$ for $T=0.3$ and
$p=0.9$. The optimal value of $q$ in this case is given by $0.85$,
whereas the weight of the excited state for the Gibbs state is $0.03$.
Thus, the optimal state $\mu$ in this case is far away from the resource-free
state.

\subsubsection{Every qubit state is optimal for some thermal noise}

\begin{figure}
\includegraphics[width=0.94\columnwidth]{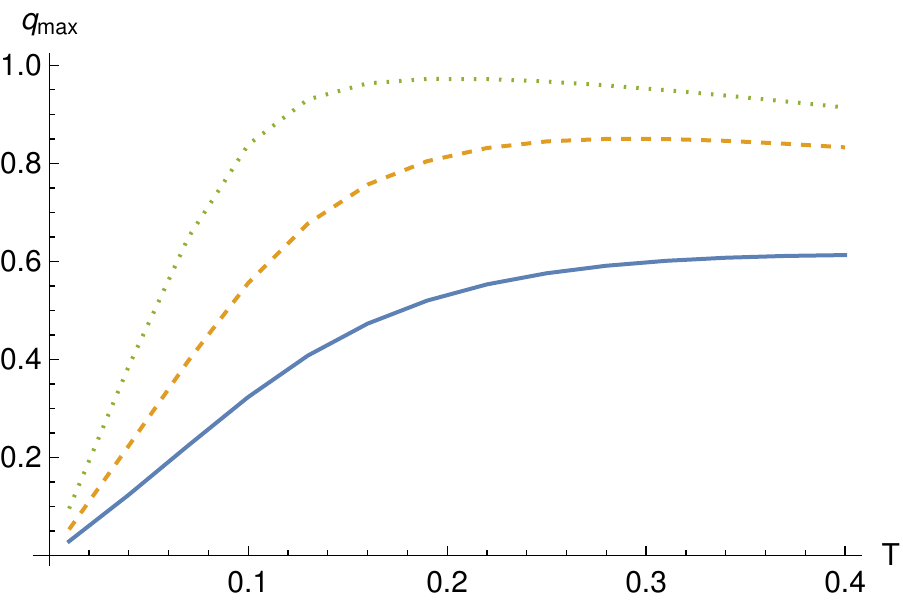}

\caption{Optimal states for thermal noise given in Eq.~(6) of the main text.
The curves show optimal parameter $q_{\max}$ {[}see Eq.~(7) of the main text{]}
as a function of temperature $T$ for $p=0.5$ (solid curve), $p=0.9$
(dashed curve) and $p=0.99$ (dotted curve).}
\label{fig:ThermalNumerics} 
\end{figure}

Here we will present numerical evidence that for every noisy qubit
state $\mu$ there exists some thermal noise $\Lambda$ such that
diluting the excited state into $\mu$ provides optimal performance.
For this, we consider noise of the form (6) of the main text,
which depends on the parameter $p$ and temperature $T$. Recalling
that the optimal state $\mu$ is diagonal in the energy eigenbasis
{[}see Eq.~(7) of the main text{]}, we numerically estimate
the optimal parameter $q$ for different values of $p$ and $T$.
The result is shown in Fig.~\ref{fig:ThermalNumerics}, where we
display the maximal value of $q$ as a function of temperature $T$
for $p=0.5$, $p=0.9$, and $p=0.99$. As is visible from the figure,
any value of $q$ is optimal for some $p$ and $T$.

\subsubsection{Reducing purity loss for depolarizing noise}
Here, we will show that in the resource theory of purity, the optimal dilution procedure to protect a pure state on a Hilbert space of dimension $d$ from depolarizing noise 
\begin{equation}
\Lambda[\rho]=p\frac{\openone}{d}+(1-p)\rho
\end{equation}
is to dilute the state as close as possible to the maximally mixed state, i.e., 
\begin{equation}
\sup_{\rho}\frac{S\left(\Lambda\left[\rho\right]||\openone/d\right)}{S\left(\rho||\openone/d\right)}=\lim_{\rho\rightarrow\id/d}\frac{S\left(\Lambda\left[\rho\right]||\openone/d\right)}{S\left(\rho||\openone/d\right)}
\end{equation}

Suppose $\rho_{t}$ is the trajectory of an initial state $\rho_{0}$
under semigroup $\Lambda_{t}$: 
\begin{equation}
\rho_{t}=\Lambda_{t}\,\rho_{0}=e^{-\gamma t}\rho_{0}+(1-e^{-\gamma t})\frac{\openone}{d}.
\end{equation}
If $\{\lambda_{i}(t)\}_{i=1}^{d}$ are eigenvalues of $\rho_{t}$,
then for $t>0$, 
\begin{multline}
\frac{d}{dt}S(\rho_{t}||\frac{1}{d}\openone)=\sum_{i=1}^{d}\dot{\lambda_{i}}(t)\log\lambda_{i}(t)=\\
-\gamma e^{-\gamma t}\sum_{i=1}^{d}(\lambda_{i}(0)-\frac{1}{d})\log\lambda_{i}(t)=\\
-\gamma\sum_{i=1}^{d}(\lambda_{i}(t)-\frac{1}{d})\log\lambda_{i}(t)=\gamma S(\rho_{t})+\frac{\gamma}{d}\log\det\rho_{t}=\\
-\gamma S(\rho_{t}||\frac{1}{d}\openone)+\gamma\log d+\frac{\gamma}{d}\log\det\rho_{t}.
\end{multline}
Because, for $t>0$, the function $\log\det\rho_{t}$ increases strictly
as $t\rightarrow\infty$, whereas at the same time the function $S(\rho_{t}||\frac{1}{d}\openone)$
strictly decreases, we conclude from the above calculation that for
$t_{1}>t_{0}>0$, 
\begin{equation}
\frac{\frac{d}{dt}S(\rho_{t}||\frac{1}{d}\openone)\Big|_{t=t_{1}}}{S(\rho_{t_{1}}||\frac{1}{d}\openone)}\,>\,\frac{\frac{d}{dt}S(\rho_{t}||\frac{1}{d}\openone)\Big|_{t=t_{0}}}{S(\rho_{t_{0}}||\frac{1}{d}\openone)}.\label{eq:relentropMonotonic}
\end{equation}
This in turn implies that the function 
\begin{equation}
f(\delta)=\frac{S(\rho_{t_{1}+\delta}||\frac{1}{d}\openone)}{S(\rho_{t_{0}+\delta}||\frac{1}{d}\openone)}
\end{equation}
is increasing for small $\delta>0$. Indeed, the fact follows directly
from \eqref{eq:relentropMonotonic}, by computing the derivative $\frac{d}{d\delta}f(\delta)\Big|_{\delta=0}$,
and from the semigroup property of the evolution $\rho_{t}$.

Setting $\Lambda=\Lambda_{t_{1}-t_{0}}$, i.e. $\Lambda\rho_{t_{0}}=\rho_{t_{1}}$,
we see that 
\begin{equation}
\frac{S(\Lambda(\rho_{t_{0}+\delta})||\frac{1}{d}\openone)}{S(\rho_{t_{0}+\delta}||\frac{1}{d}\openone)}>\frac{S(\Lambda(\rho_{t_{0}})||\frac{1}{d}\openone)}{S(\rho_{t_{0}}||\frac{1}{d}\openone)}
\end{equation}
for any $t_{0}>0$, small $\delta>0$, and any initial state $\rho_{0}$.
Hence the supremum 
\begin{equation}
\sup\limits _{\rho}\frac{S(\Lambda(\rho)||\frac{1}{d}\openone)}{S(\rho_{}||\frac{1}{d}\openone)}
\end{equation}
is attained for $\rho\rightarrow\frac{1}{d}\openone$.

\subsubsection{Correlations in the resource theory of purity}

We will now consider unital single-qubit noise. As discussed in the main text, it is possible to reduce the loss of purity under
unital noise by diluting the single-qubit pure states into noisy states
of a single qubit. However, it remained unclear if the procedure can
be improved by establishing correlations between the qubits, i.e,
whether creating a correlated pure $k$-qubit state $\ket{\psi_{k}}$
leads to a better performance. Here, we will prove that this is not
the case: correlations are not useful in this procedure when considering
pure states of $k$ qubits. For this, note that the figure of merit
in this case is given by 
\begin{equation}
\frac{S(\Lambda^{\otimes k}[\psi_{k}]||\openone_{2^{k}}/2^{k})}{S(\psi_{k}||\openone_{2^{k}}/2^{k})}=1-\frac{S(\Lambda^{\otimes k}[\psi_{k}])}{k}.\label{eq:PurityCorrelations}
\end{equation}
We are interested in the maximum of this quantity, maximized over
all pure $k$-qubit states $\ket{\psi_{k}}$. Note that maximizing
the right-hand side of Eq.~(\ref{eq:PurityCorrelations}) corresponds
to minimizing the output entropy of $\Lambda^{\otimes k}$. Recall
that the minimal output entropy of $\Lambda_{1}\otimes\Lambda_{2}$
is additive if $\Lambda_{1}$ is a unital qubit channel~\cite{King2022}.
Thus, we conclude that for any $k$-qubit pure state $\ket{\psi_{k}}$
there exists a pure qubit state $\ket{\phi_{1}}$ and a pure $k-1$-qubit
state $\ket{\phi_{k-1}}$ such that 
\begin{equation}
S(\Lambda^{\otimes k}[\psi_{k}])\geq S(\Lambda[\phi_{1}])+S(\Lambda^{\otimes k-1}[\phi_{k-1}])=S(\Lambda^{\otimes k}[\phi_{1}\otimes\phi_{k-1}]).
\end{equation}
Iterating this procedure, we see that for any $\ket{\psi_{k}}$ there
exists a pure single-qubit state $\ket{\phi}$ such that 
\begin{equation}
S(\Lambda^{\otimes k}[\psi_{k}])\geq kS(\Lambda[\phi]).
\end{equation}
This proves that the maximum of Eq.~(\ref{eq:PurityCorrelations})
(when maximized over pure $k$-qubit states) is achieved on product
states.

\subsection{Strategies beyond dilution}
\subsubsection{Quantum Error Correction}
\paragraph{Resource theory of entanglement: phase-flip channel}
\label{app:three_qubit_phase_flip}
In the main text, we proposed a way to reduce the loss of entanglement through a phase-flip channel by diluting (before the noise) the $n$ singlets into $n/S(\psi^A)$ copies of the state $\ket{\psi}=\cos(\alpha) \ket{00}+\sin(\alpha) \ket{11}$. This process can be done with LOCC operations (and $1/S(\psi^A)-1$ additional qubit in each laboratory per initial singlet). Another strategy could be to use additional qubits, not to dilute the singlet, but to encode it with an error-correction code before the noise acts. Once the noise has acted, the error-correction code would detect which error the noise introduced, correct it and decode the state. The resulting state would then be used as an input to the final distillation protocol. 

Because the noise model is a phase-flip channel, it is natural to use the three qubit phase-flip code as an error correction code \cite{nielsen2002quantum}. We now describe step-by-step how we use it. 

\textbf{First step:} We encode the singlet with two additional qubits initialized in $\ket{+}$ in Bob's lab. Before the noise acts, a sequence of cNOTs in Bob's lab allows to encode the singlet in a state that we call $\ket{\psi_{\text{enc}}}$:
\begin{align}
    (\ket{+_A+_B}+\ket{-_A-_B}) \otimes  \ket{+_B+_B} \to \nonumber\\
   \to \ket{\psi_{\text{enc}}} \equiv \ket{+_A (+++)_B}+\ket{-_A (---)_B}.
\end{align}
In this last equation, we have put indices $A$ and $B$ to refer to which laboratory each qubit is in. The notation $\ket{(+++)_B}$ means three qubit in Bob's lab in the $\ket{+}$ state. We also used the fact that $\ket{00}+\ket{11}=\ket{++}+\ket{--}$ to write the initial singlet as $\ket{+_A+_B}+\ket{-_A-_B}$. 

\textbf{Second step:} The phase-flip channel acts on each qubit in Bob's laboratory. Rewriting the phase-flip channel as:
\begin{align}
    &\Lambda(\rho)=(1-p) \rho + p Z \rho Z \notag \\
    & p \equiv \frac{1}{2}(1-\sqrt{1-\lambda}),
\end{align}
we can show that the state after the noise has acted reads:
\begin{widetext}
\begin{align}
    &\openone \otimes \Lambda^{\otimes 3}(\ket{\psi_{\text{enc}}} \bra{\psi_{\text{enc}}}) \notag \\
    &= (1-p)^3 \ket{\psi_{\text{enc}}} \bra{\psi_{\text{enc}}} + p(1-p)^2 \sum_{i=2}^4 Z_i \ket{\psi_{\text{enc}}} \bra{\psi_{\text{enc}}} Z_i+ p^2(1-p) \sum_{2 \leq i<j\leq 4} Z_i Z_j \ket{\psi_{\text{enc}}} \bra{\psi_{\text{enc}}} Z_i Z_j  + p^3 Z_2 Z_3 Z_4 \ket{\psi_{\text{enc}}} \bra{\psi_{\text{enc}}} Z_2 Z_3 Z_4.
    \label{eq:rho_noise}
\end{align}
\end{widetext}
In Eq.~\eqref{eq:rho_noise}, we use the notation $Z_i$ to indicate a Pauli $Z$ operator applied on the $i$'th qubit (and the identity on the other qubits).

\textbf{Third step:} We measure the observables $X_2 X_3$ (outcome stored in a bit $x_{23}$) and $X_3 X_4$ (outcome stored in a bit $x_{34}$). These measurements are local in Bob's lab. We then apply a unitary $R$ ("recovery") on the qubits in Bob's lab, following the rule provided in Table \ref{table:recovery}.
\begin{table} [b]
\centering
\begin{tabular}{|c|c|}
    \hline
     $(x_{23},x_{34})$ &  $R$ \\
     \hline
     $(+1,+1)$ & $I$ \\
     \hline
     $(+1,-1)$ & $Z_4$ \\
     \hline
     $(-1,+1)$ & $Z_2$ \\
     \hline
     $(-1,-1)$ & $Z_3$ \\
     \hline
\end{tabular}
\caption{Recovery operation $R$ to be applied in order to fix the phase-flip errors in \eqref{eq:rho_noise}.}
\label{table:recovery}
\end{table}
The density matrix after this recovery will have the expression:
\begin{align}
    \rho_{\text{recovery}}&= \left( (1-p)^3 +3p(1-p)^2 \right) \ket{\psi_{\text{enc}}} \bra{\psi_{\text{enc}}} \notag \\
    &+ \left( 3 p^2(1-p)+p^3 \right) Z_2 Z_3 Z_4\ket{\psi_{\text{enc}}} \bra{\psi_{\text{enc}}} Z_2 Z_3 Z_4.
    \label{eq:post_recovery}
\end{align}
While a brute force calculation would show that \eqref{eq:post_recovery} is satisfied, we can understand the protocol as follows. The first order terms in $p$ in \eqref{eq:rho_noise} correspond to the application of one Pauli $Z$ operator (on either of the qubits $2$, $3$ or $4$). This Pauli operator will change the "parity" between some of these qubits. For instance, $Z_2 \ket{\psi_{\text{enc}}}=\ket{+_A (-++)_B}+\ket{-_A (+--)_B}$: the qubits $3$ and $4$ have the same parity, but not the qubits $2$ and $3$. This is detected by the measurement, which would provide the outcome $(x_{23}, x_{34})=(-1,+1)$. The error is then inverted by applying the unitary $Z_2$, giving back the original encoded state $\ket{\psi_{\text{enc}}}$. Overall, all the events of order $p$ in \eqref{eq:rho_noise} can be perfectly corrected, following the recovery described in Figure \ref{table:recovery}. However, any event of order $p^2$ or $p^3$ (which is less likely to occur given the fact $p<1/2$) would lead to an erroneous correction. For instance, a pair of phase-flip, $Z_3 Z_4$, occurring with a probability $p^2(1-p)$, would also provide a measurement outcome $(x_{23}, x_{34})=(-1,+1)$, suggesting that the recovery $Z_2$ should be applied. In such a case the final state would be $Z_2 Z_3 Z_4 \ket{\psi_{\text{enc}}} \neq \ket{\psi_{\text{enc}}}$. In the end, the state in Eq.~\eqref{eq:post_recovery} is obtained.

\textbf{Fourth step:} We decode the state by applying the reverse operation from the encoding, and we discard the additional qubits we used. We would then obtain the final state:
\begin{align}
    \rho_{\text{decoded}}&= \left( (1-p)^3 +3p(1-p)^2 \right) \ket{\psi^-} \bra{\psi^-} \notag \\
    &+ \left( 3 p^2(1-p)+p^3 \right) Z_2 \ket{\psi^-} \bra{\psi^-} Z_2.
    \label{eq:decoded}
\end{align}
It is this state that we will distill in order to obtain perfect singlets.

\begin{figure}[b]
    \centering
    \includegraphics[width=1\columnwidth]{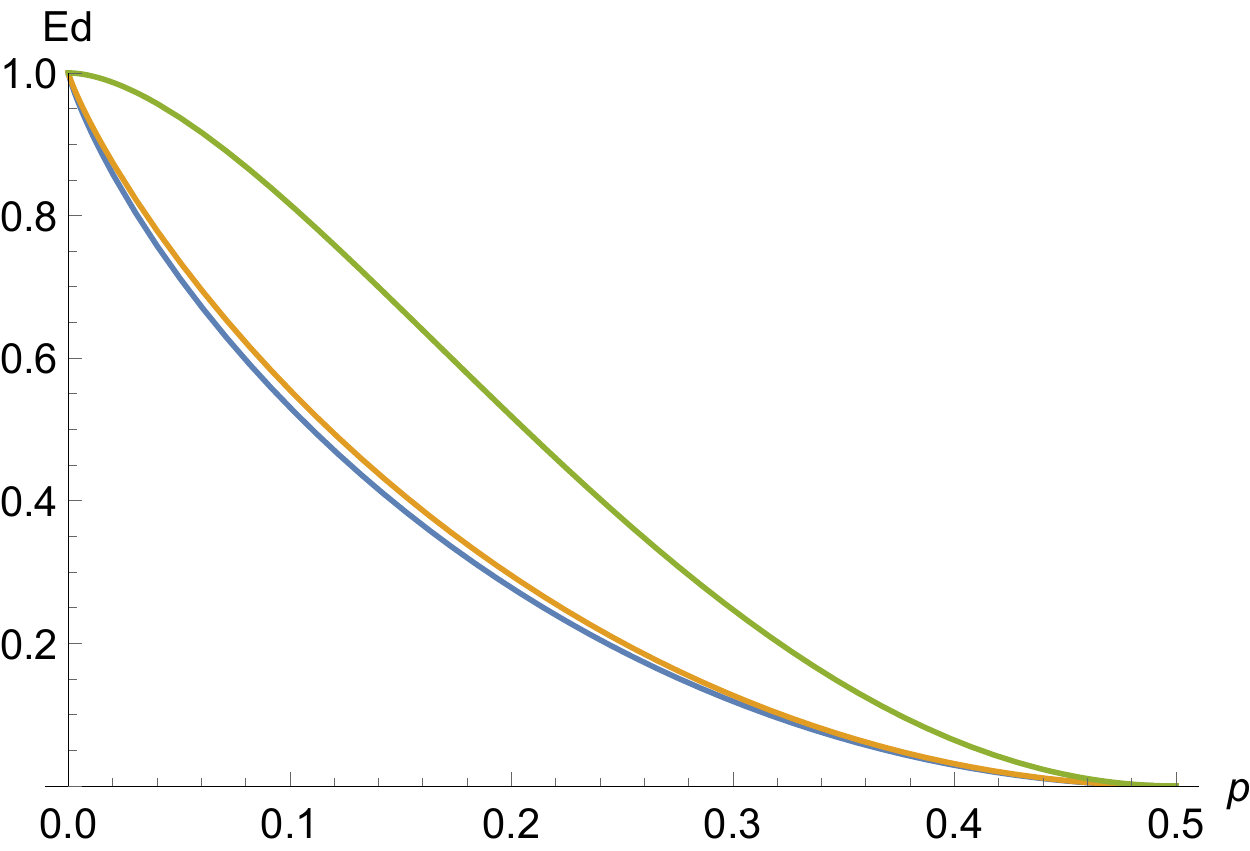}
    \caption{Comparison of $E_d^{Dil}(p)$ for $\alpha=0.25$ s.t. $1/S(\Psi^A)=3$ (in orange), $E_d^{QEC}(p)$ (in green) and $E_d(p)$ when no kind of pre-processing is applied before the noise (in blue). For every value of $p$ dilution is an advantageous pre-processing scheme, but it is never better than error correction.}
    \label{Dil_ErrCorr_plot}
\end{figure}
\textbf{Comparing dilution and error-correction strategy:} In the calculation done above, we used two extra qubits in Bob's laboratory to perform error correction. It allowed to make $\rho_{\text{decoded}}$ noiseless up to order $p$. If Bob had no limit in the number of extra qubits he could use, he could generalize the calculations shown above by using a phase flip code using $2t+1$ qubits instead of $3$ \cite{wootton2018repetition}. Such code would be able to detect (and correct) errors up to order $t$. The encoded state, before the noise, would be generalized as:
\begin{align}
\ket{\psi_{\text{enc}}^{(2t+1)}} \equiv \ket{+_A}\ket{+_B}^{\otimes (2t+1)}+\ket{-_A}\ket{-_B}^{\otimes (2t+1)}.
\end{align}
The post-decoded state (after the noise has acted and error correction has been applied, following a generalization of the protocol described, to the $2t+1$ qubit phase-flip code), discarding the $2t$ extra ancillas would have the form:  \begin{align}
    \rho_{\text{decoded}}&= \left( 1-P(p,t) \right) \ket{\psi^-} \bra{\psi^-} \notag \\
    &+ P(p,t) Z_2 \ket{\psi^-} \bra{\psi^-} Z_2,
\end{align}
with 
\begin{align}
    P(p,t)=\sum_{k=t+1}^{2t+1} \binom{2t+1}{k}p^k (1-p)^{2t+1-k}.
\end{align}
Physically, $P(p,t)$ corresponds to the probability that more than $t+1$ errors occurred. For any $p<1/2$, one can check numerically that $\lim_{t \to +\infty} P(p,t)=0$, meaning that using enough extra ancillary qubits, error correction could perfectly fix the errors.

This is why, in order to make a fair comparison, we should only allow using the same number of additional qubit per initial singlet both in the dilution and in the error-correction protocols. In what follows, we assume that a three-qubit phase-flip code is used, meaning that the dilution protocol must satisfy: $1/S(\psi^A)=3$. This means that each of the parties, Alice and Bob, are allowed to use two ancillary qubits per each shared singlet. Because $\rho_{\text{decoded}}$ is maximally correlated, the distillation rate for error correction can be evaluated as:
\begin{align}
    E_d^{QEC}(p)=S(Tr_B(\rho_{\text{decoded}}))-S(\rho_{\text{decoded}})=\nonumber\\=1-h((1-p)^3+3p(1-p)^2)
    \label{eq:Ed_QEC}
\end{align}
where $S(\rho)$ corresponds to the Von Neuman entropy contained in $\rho$. Using \eqref{eq:distillable_entanglement_advantagedilution}, and replacing $\lambda=1-(1-2p)^2$, the distillation rate for the dilution strategy reads:
\begin{align}
    &E_d^{Dil}(p)=1-\frac{h\left(\frac{1}{2}\left(1+\sqrt{1-2p(1-p)(1-\cos{4\alpha})}\right)\right)}{h(\cos^2{\alpha})}, \notag \\
    &\alpha=0.25
    \label{eq:Ed_dil}
\end{align}

In Figure \ref{Dil_ErrCorr_plot}, we see that error-correction outperforms dilution for all values of $p$.
\paragraph{Resource theory of entanglement: general Pauli channel}
A natural question one can ask is if for more general noise models, dilution could outperform quantum error correction. Here, we study this question by considering a general Pauli noise channel, $\Lambda_{\mathbf{p}}(\rho) \equiv \sum_{i=0}^3 p_i \sigma_i \rho \sigma_i$, where $\mathbf{p}=(p_0,p_1,p_2,p_3)$ is a vector containing probabilities ($p_0=1-(p_1+p_2+p_3)$). This noise channel is applied to each qubit in Bob's laboratory. We will show that performing dilution before the noise acts does not outperform quantum error correction, under the constraint of using $2$ additional ancilla qubits per singlet.

To begin our explanations, we assume that we use the three-qubit phase flip code. Hence, we apply the exact protocol described in Section \ref{app:three_qubit_phase_flip}, but we replace the phase-flip noise model with the Pauli noise channel $\Lambda_{\mathbf{p}}(\rho)$. In such a case, one can show that the density matrix of the decoded state, written in the basis $(\ket{++},\ket{+-},\ket{-+},\ket{++})$, once the two extra ancillas used to encode the state have been traced out, reads:
\begin{widetext}
\begin{equation}\label{theta}
\rho^{\text{Pauli}}_{\text{decoded}}(\mathbf{p}) =
\!\begin{aligned}
&
\left(\begin{matrix}
  \frac{1}{2}(-1+p_2+p_3)^2(1+2p_3+2p_2) & 0\\
  0 & -\frac{1}{2}(p_2+p_3)^2(-3+2p_2+2p_3)\\
  0 & -\frac{1}{2}(-p_2+p_3)^2(-3+6p_1+4p_2+2p_3)\\
  \frac{1}{2}(-1+2p_1+p_2+p_3)^2(1-2p_1-4p_2+2p_3)& 0\\
\end{matrix}\right.\\
&\qquad\qquad
\left.\begin{matrix}
  0 &\frac{1}{2}(-1+2p_1+p_2+p_3)^2(1-2p_1-4p_2+2p_3) \\
  -\frac{1}{2}(-p_2+p_3)^2(-3+6p_1+4p_2+2p_3)& 0\\
   -\frac{1}{2}(p_2+p_3)^2(-3+2p_2+2p_3)& 0\\
    0 & \frac{1}{2}(-1+p_2+p_3)^2(1+2p_2+2p_3)\\
\end{matrix}\right)
\end{aligned}
\end{equation}    
\end{widetext}
Now, in general, the phase-flip code will only be efficient in the case $p_3 \gg p_2,p_1$. In order to design an efficient protocol, we can apply a unitary $V$ right before and after the noise acts in such a manner that: $V \Lambda_{\mathbf{p}} V^{\dagger}=\Lambda_{\mathbf{p}^V}$, where $\mathbf{p}^V=(p^V_1,p^V_2,p^V_3)$ is a permutation of the elements of the vector $\mathbf{p}=(p_1,p_2,p_3)$ such that $p^V_3=\max(p_1,p_2,p_3)$. To simplify our numerical calculations, we added the constraint that $\mathbf{p}^V$ is deduced from $\mathbf{p}$ through a \textit{cyclic} permutation of its coefficients. To be more formal, defining $k \equiv Arg[\max_{i>0} p_i]$, we have $\mathbf{p}^V=(p_{k+1},p_{k+2},p_k)$, where the additions $k+1$ and $k+2$ are performed modulo $3$. For instance, if $k=2$, we would have $\mathbf{p}^V=(p_{3},p_{1},p_2)$. We should notice that the error-correction strategy we are using is not the optimal one in general for at least two reasons. First, we could use an error-correction code able to correct against an arbitrary single-qubit error, like the 5 qubit code \cite{simakov2022simulation}. Then, we decided to restrict ourselves (for simplicity) to a \textit{cyclic} permutation in the choice of $\mathbf{p}^V$ (which might not be the optimal choice). In practice, applying the unitary $V$ is equivalent to evaluate $\rho^{\text{Pauli}}_{\text{decoded}}$ where the coefficients $(p_1,p_2,p_3)$ in its expression are permuted to $(p_{k+1},p_{k+2},p_k)$. This is the input density matrix we use in order to compute the distillable entanglement for our error-correction scheme. Recall that for any bipartite state $\rho^{AB}$ the distillable entanglement is bounded as~\cite{Devetak2005}:
\begin{equation}
    E_d(\rho^{AB}) \geq S(\rho^A)-S(\rho^{AB}), \label{eq:HashingBound}
\end{equation}
and equality is achieved on maximally correlated states, i.e., states of the form $\rho = \sum_{i,j} \rho_{ij} \ket{ii}\!\bra{jj}$. Denoting with $E^{QEC}_d(\mathbf{p})$ the distillable entanglement of the state $\rho^{\text{Pauli}}_{\text{decoded}}$ we thus have
\begin{equation}
    E^{QEC}_d(\mathbf{p}) \geq S(\mathrm{Tr}_B(\rho^{\text{Pauli}}_{\text{decoded}}(\mathbf{p}^V)))-S(\rho^{\text{Pauli}}_{\text{decoded}}(\mathbf{p}^V)).
\end{equation}

We compare this strategy with the best dilution protocol we can do, assuming only two extra ancillas are used per singlet. This optimal dilution protocol is defined as follow. We first dilute the singlets in $3$ copies of less entangled states per unit of initial ones. The diluted states will be of the form $\ket{\Psi}=\cos\alpha\ket{00}+\sin{\alpha}\ket{11}$, with $\alpha \approx 0.25$ s.t. $1/S(\Psi^A)=3$, because we want to compare to the 3 qubit error correcting code. We then rotate the local basis of Bob by the application of a single-qubit unitary $U$ in his lab. Bob has total control over this unitary and he can choose the one which suits best the procedure. After this pre-processing the Pauli noise acts locally on each copy of Bob's particles
\begin{equation}    \rho=\openone\otimes\Lambda_p(\openone\otimes U\ket{\Psi}\bra{\Psi}\openone\otimes U^{\dagger})
\end{equation}
The total rate of distillable singlets from this procedure will be
\begin{equation}    E_d^{dil}=\frac{R(\rho\rightarrow\Phi^+)}{S(\Psi^A)}
\end{equation}
Since $R(\rho\rightarrow\Phi^+)$ corresponds to the distillable entanglement of $\rho$, it can be lower bounded by Eq.~(\ref{eq:HashingBound}), with the equality achieved when $\rho$ is maximally correlated.
So we can bound our distillable entanglement by
\begin{equation}
    E_d^{dil}\geq\frac{S(\rho^B)-S(\rho)}{S(\Psi^B)}=BE_d^{dil}
\end{equation}
where we choose to trace out Alice's system because Bob's noisy state will have a higher entropy, i.e. $S(\rho^B)\geq S(\rho^A)=S(\Psi^A)=S(\Psi^B)$.
We are in particular interested in a dilution protocol that maximizes this bound on the distillable entanglement, the maximization being performed over $U$. We performed numerical calculations by sampling over a wide range of $\mathbf{p}$, and we compared the strategies based on the optimal dilution protocol, error-correction, and the strategy that "does nothing" (i.e. no pre-processing is performed before the noise). For all the points we calculated, we realized that as soon as the lower bound on distillable entanglement was positive, the optimal strategy was either error-correction or the "do nothing" strategy: the dilution never outperformed both. We note that these results refer to a comparison of two lower bounds, which means that a rigorous conclusion about the advantage of quantum error correction over the dilution strategy cannot be drawn at the moment.

\subsubsection{More general strategies}
For simplicity, we will focus on the resource
theory of entanglement, but the presented ideas are also applicable
to other resource theories. In particular, we assume that Alice and
Bob shares $n$ singlets initially, aiming to protect them from local
qubit noise on Bob's side. The most general strategy is to first convert
the $n$ singlets into $2m$-qubit state $\rho_{2m}$, where Alice
and Bob each hold $m$ of the qubits. After the action of noise, Alice
and Bob end up sharing the state $\openone^{\otimes m}\otimes\Lambda^{\otimes m}[\rho_{2m}]$.
In the final step, they perform local operations and classical communication,
aiming to convert the state $\openone^{\otimes m}\otimes\Lambda^{\otimes m}[\rho_{2m}]$
into $k$ singlets. The maximal possible singlet rate $k/n$ achievable
in this process can be seen as a figure of merit in this task, which
we term \emph{entanglement protection rate}.

Here we provide a formal definition of the entanglement protection
rate for any type of quantum noise $\Lambda$ acting on a Hilbert
space of dimension $d$. Let $\Phi_{1}$ be an LOCC protocol which
takes $2n$ qubits as an input and gives $2m$ quantum systems of
dimension $d$ at the output, and in both cases one half of the
system belongs to Alice and Bob, respectively. Similarly, $\Phi_{2}$
is an LOCC protocol which acts on $2m$ quantum systems of dimension
$d$, producing a quantum state of $2k$ qubits at the output. Recalling
the definition of quantum fidelity $F(\rho,\sigma)=\mathrm{Tr}[\sqrt{\rho^{1/2}\sigma\rho^{1/2}}]$,
we are now ready to define the single-shot fidelity for entanglement
protection: 
\begin{equation}
F(\Lambda,n,k)\!=\!\sup_{\Phi_{1},\Phi_{2}}\!F\!\left(\Phi_{2}\!\left[\!\openone^{\otimes m}\!\otimes\!\Lambda^{\otimes m}\left(\Phi_{1}\!\left[\ket{\phi^{+}}\!\bra{\phi^{+}}^{\otimes n}\right]\right)\right]\!,\!\ket{\phi^{+}}\!\bra{\phi^{+}}^{\otimes k}\right).
\end{equation}
With this, we define the entanglement protection rate of $\Lambda$
as follows: 
\begin{equation}
E_{\mathrm{p}}\left(\Lambda\right)=\sup\left\{ r:\lim_{n\rightarrow\infty}F\left(\Lambda,n,\left\lfloor rn\right\rfloor \right)=1\right\} .
\end{equation}

Clearly, the entanglement protection rate is lower bounded by the
rate achieved via entanglement dilution procedure described above
in this article: 
\begin{equation}
E_{\mathrm{p}}\left(\Lambda\right)\geq\sup_{\rho}\frac{E_{\mathrm{d}}\left(\openone\otimes\Lambda\left[\rho\right]\right)}{E_{\mathrm{c}}\left(\rho\right)},
\end{equation}
where the supremum is taken over all bipartite states $\rho$. As discussed above, for some types of noise a better performance can be achieved by using strategies based on quantum error correction.

\end{document}